\documentclass[11pt,letterpaper]{article}

\pdfoutput=1  

\usepackage{palatino}
\usepackage[left=1in,top=1in,right=1in,bottom=1in]{geometry}
\usepackage{amsmath}
\usepackage{amssymb}
\usepackage{amsbsy}
\usepackage{amsthm}
\usepackage{epsfig}
\usepackage{clrscode}
\usepackage{setspace}
\usepackage{multirow}
\usepackage{url}
\usepackage{paralist}
\usepackage{wrapfig,boxedminipage}
\usepackage{graphicx}
\usepackage{subcaption}
\usepackage{float}
\usepackage{xcolor}
\usepackage{pgfplotstable}
\usepackage{hyperref}
\usepackage{enumitem} 
\usepackage{adjustbox}
\usepackage{booktabs}
\usepackage{csvsimple}
\usepackage{tikz}
\usepackage{dcolumn}
\usepackage{physics}  

\usepackage{enumitem}
\setitemize{noitemsep,topsep=0pt,parsep=0pt,partopsep=0pt}

\usepackage{setspace} 
\usepackage{indentfirst}

\usepackage{graphicx}
\newcommand{\indep}{\rotatebox[origin=c]{90}{$\models$}}

\usepackage{bibentry}

\usepackage{tikz}
\usetikzlibrary{shapes, calc, shapes, arrows, snakes}
\usetikzlibrary{positioning}
\newdimen\nodeDist
\tikzset{dist1/.style={path picture= {
    \begin{scope}[x=1pt,y=10pt]
      \draw plot[domain=-6:6] (\x,{\x/24});
    \end{scope}
    }
  }
}
\tikzset{dist2/.style={path picture= {
    \begin{scope}[x=1pt,y=10pt]
      \draw plot[domain=-6:6] (\x,{0.01*(\x-2)^2-0.4});
    \end{scope}
    }
  }
}
\tikzset{dist3/.style={path picture= {
    \begin{scope}[x=1pt,y=10pt]
      \draw plot[domain=-6:6] (\x,{0.85/(1 + exp(\x))-0.5});
    \end{scope}
    }
  }
}
\tikzstyle{f1}=[draw,circle,minimum size=25pt,inner sep=0pt,dist1]
\tikzstyle{f2}=[draw,circle,minimum size=25pt,inner sep=0pt,dist2]
\tikzstyle{f3}=[draw,circle,minimum size=25pt,inner sep=0pt,dist3]

\usepackage[round]{natbib}

\newcommand{\nc}{\newcommand}

\nc{\bX}{{\bf X}}
\nc{\lhat}[1][i]{\hat\lambda_{#1}^{-1(g)}}
\nc{\what}[1][j]{\hat\omega_{#1}^{-1(g)}}
\nc{\Li}{\hat\Lambda^{-1(g)}}
\nc{\Oi}{\hat\Omega^{-1(g)}}
\nc{\diag}[1]{\text{diag}\left(#1\right)}
\nc{\Siginv}{\Sigma^{-1}}
\nc{\Ominv}{\Omega^{-1}}
\nc{\bone}{{\bf 1}}

\newcommand{\x}{\mathbf{x}}

\newcommand{\beq}{\begin{equation}}
\newcommand{\eeq}{\end{equation}}

\newcommand{\ben}{\begin{enumerate}}
\newcommand{\een}{\end{enumerate}}

\providecommand{\keywords}[1]{\textbf{\textit{Keywords and phrases: }} #1}

\newcommand{\methN}{\text{Targeted Smooth Bayesian Causal Forests}} 
\newcommand{\methA}{\text{tsBCF}} 

\begin{document}

\newtheorem{assump}{Assumption}

\title{\methN{}: \\
An analysis of heterogeneous treatment effects for simultaneous versus interval medical abortion regimens over gestation}
\author{Jennifer E.~Starling$^\dag$\footnote{Corresponding author: jstarling@utexas.edu} \and
		Jared S. Murray$^{\dag, \ddag}$ \and 
		Patricia A. Lohr$^{\S}$ \and
		Abigail R.A. Aiken$^{||}$ \and
		Carlos M.~Carvalho$^{\dag, \ddag}$ \and
		James G.~Scott$^{\dag, \ddag}$
		}

\date{%
    $^\dag$Department of Statistics and Data Sciences, \\
	$^\ddag$McCombs School of Business, and \\
	$^{||}$Lyndon B. Johnson School of Public Affairs, \\
	The University of Texas at Austin, Austin, TX \\ ~\\
    $^\S$British Pregnancy Advisory Service, Stratford-upon-Avon, UK\\[2ex]%
}

\maketitle

\begin{abstract}


We introduce \methN{} (\methA{}), a nonparametric Bayesian approach for estimating heterogeneous treatment effects which vary smoothly over a single covariate in the observational data setting.  
The \methA{} method induces smoothness by parameterizing terminal tree nodes with smooth functions, and allows for separate regularization of treatment effects versus prognostic effect of control covariates. 
Smoothing parameters for prognostic and treatment effects can be chosen to reflect prior knowledge or tuned in a data-dependent way.

We use tsBCF to analyze a new clinical protocol for early medical abortion.  
Our aim is to assess relative effectiveness of simultaneous versus interval administration of mifepristone and misoprostol over the first nine weeks of gestation.  
The model reflects our expectation that the relative effectiveness varies smoothly over gestation, but not necessarily over other covariates.
We demonstrate the performance of the \methA{} method on benchmarking experiments.  
Software for tsBCF is available at \url{https://github.com/jestarling/tsbcf/}. 
\end{abstract}


\newpage
\onehalfspace

\begin{centering}\section{Introduction} \end{centering}

We analyze data on a new clinical protocol for early medical abortion (EMA) using data from British Pregnancy Advisory Service (BPAS), a non-profit abortion provider in the United Kingdom.  
Early medical abortion uses a combination of mifepristone and misoprostol, and is available through 63 days of gestation in Britain.  
The current standard of care is an ``interval'' protocol, where 200 mg oral mifepristone is administered in a clinic setting, followed by administration of 800 micrograms vaginal misoprostol during a second clinic visit 24--48 hours later \citep{rcog2011}.  
BPAS recently began offering the option of a ``simultaneous'' protocol, where patients receive both medications in a single clinic visit, along with instructions about what to do at home to ensure safety and efficacy.

Past research has shown that patients prefer home use of mifepristone and misoprostol over the clinic setting \citep{gold2015,ngo2011}.  
When these medications are restricted to the clinical setting, as in the UK \citep{abortionact1967}, patients overwhelmingly prefer the simultaneous protocol \citep{lohr2018}.
A second clinic visit imposes financial and logistical burdens on access to care, particularly for patients from disadvantaged backgrounds or regions without a local clinic, who may incur significant expenses for even a single clinic visit.
Previous research also found that on average, the simultaneous protocol is nearly (97\%) as effective as the interval protocol \citep{li_concurrent_2006, goel_simultaneous_2011}.  
This finding raises questions about whether the small decrease in average efficacy may be magnified into a much larger gap at later gestational ages (in particular, 7--9 weeks).
This is the key scientific question that we aim to address.

Answering this question using BPAS's data is complicated by the fact that patients self-select into the simultaneous or interval protocol.  
This introduces the possibility of bias in our estimation of the relative efficacy of the two protocols across gestational age.  
In an earlier paper, \citet{lohr2018} aimed to address this using logistic regression with a propensity score adjustment.  
They confirmed the 97\% average efficacy gap found in previous research, and did not find significant evidence supporting decrease in relative efficacy over gestation.  
There are three issues with the approach taken in \citet{lohr2018}.  
First, assuming a simple parametric form for the probability of successful EMA given treatment and covariates opens the possibility of bias due to model mis-specification.  
Second, their analysis treated gestational age -- the dominant predictor of a successful EMA -- as a categorical variable, discretizing into three gestational age ranges. 
There are strong reasons to believe that the probability of a successful EMA varies smoothly over gestation.  
Third, their approach estimates average relative effectiveness of the simultaneous protocol over gestation, but does not address subgroups.  
A key clinical question is whether there are subgroups of patients for whom the efficacy gap is much wider than average, especially at higher gestational ages.  
Identifying these subgroups would have profound implications for the way that BPAS, and indeed all family-planning clinics, counsel their patients.

Our goal is to propose a model capable of addressing all of these shortcomings.  
Our model, called Targeted Smooth Bayesian Causal Forests (tsBCF),  is nonparametric and so avoids the potential for biases that arise in assuming specific functional forms for the causal estimand.  
It is capable of detecting heterogeneous treatment effects should they exist.  
Finally, it allows for smoothness over a single covariate, addressing a key statistical issue that arises in virtually all pregnancy-related research: that most outcomes vary smoothly with gestational age. 

We validate previous findings \citep{lohr2018} that on average, we do not see a marked decrease in efficacy as gestational age increases.  
However, we do find modest decrease in the 7--9 week range (from 0.972 at 6 weeks, to 0.955 at 7 weeks, to 0.948 at 9 weeks).  
We identify a more pronounced drop in efficacy at later gestations for patients age 29 and older, particularly for those who have given birth previously.
While relative effectiveness for these patients is still over 90\% in the 7--9 week range, clinicians may wish to counsel their patients accordingly.

In Section \ref{sec-ema}, we provide clinical background and an overview of the dataset.  
Section \ref{sec-tsBCF} presents the \methA{} model and reviews relevant literature. 
Section \ref{sec-results} gives results for the early medical abortion analysis.    
Section \ref{sec-sims} details a simulation study showing that tsBCF maintains accuracy and nominal coverage in recovering heterogeneous relative effectiveness for several clinically relevant scenarios.  
Section \ref{sec-discussion} concludes with discussion.  
The Appendix provides additional detail on fitting the \methA{} model.  \\

\begin{centering}\section{Early Medical Abortion Data} \label{sec-ema} \end{centering}

The dataset consists of early medical abortions provided at BPAS clinics from May 1, 2015 to April 30, 2016.  
Data was collected from BPAS's electronic booking and invoicing system, which contained records of services provided to clients, including selected demographic and clinical characteristics.  
These data were entered by telephone operators at British Pregnancy Advisory Service's telephone contact center; details were then validated by clinicians at both in-person consultations and treatment appointments. 
When possible, hospital discharge summaries or documents from general practitioners were obtained to confirm outcomes.  
Staff cross-checked the booking and invoicing system for any appointments with British Pregnancy Advisory Services after the date of treatment, and hand-checked medical records if a continuing pregnancy or incomplete abortion was recorded in the complications database.  
Complications and adverse outcomes were identified either during post-treatment follow-up visits, by patients themselves, or by notifications from other providers.
This study was approved and exempted from full human subjects review by British Pregnancy Advisory Services and The University of Texas at Austin since all data were pre-existing and were provided in a fully de-identified format.

The dataset consists of 28,895 independent patient records; each record is a pregnancy between 4.5 weeks (32 days) and 9 weeks (63 days) gestation, as determined by abdominal or vaginal ultrasonography, where the patient sought a medical abortion and had no contraindications.
Patients are typically not aware of pregnancy prior to 4 weeks, and only 12 patients obtained early medical abortions below 4.5 weeks; our analytics sample begins at 4.5 weeks.
Early medical abortion is unavailable after 9 weeks.

Patients were offered the choice between the interval regimen, where 200 mg oral mifepristone was administered in the first clinic visit followed by 800 micrograms vaginal misoprostol in a second clinic visit 24--72 hours later, and the simultaneous regimen, where the medications were administered in a single clinic visit no more than 15 minutes apart. 
Patients received counseling from their provider before self-selecting to a regimen.
Our analytic sample consists of patients who chose simultaneous dosing or a 24--48 hour interval between medications.

Our binary response is successful early medical abortion outcome, defined as complete evacuation of uterine contents without surgical intervention and without continuing pregnancy \citep{creinin2016} .
Patients could choose to return two weeks post-treatment for a vaginal ultrasound, or could use a low-sensitivity urine pregnancy test (detection limit 1,000 international units human chorionic gonadotrophin) and symptom checklist to self-report the outcome of the abortion \citep{cameron2015}. 
Patients could schedule a clinic visit at any time to address concerns or symptoms of a possible complication, including continuing pregnancy.  
Patients diagnosed at a follow-up visit with a retained nonviable sac or embryo were offered the choice of another 800 micrograms vaginal mifepristone or surgical evacuation, and patients diagnosed with continued pregnancy were offered surgical evacuation; all of these are considered unsuccessful outcomes.\\

\begin{table}[ht] 
\centering
\scalebox{0.65}{
\begin{tabular}{lrrrr}
 Characteristic & (N = 28,895) & Interval: (N = 4,354) & Simultaneous: (N = 24,541) & P-value \\ 
  \hline
Gestational age in weeks, n (\%) &  &  &  & $<$ 0.0001 \\ 
   \hline
4.5 & 407 (1.41) & 20 (0.46) & 387 (1.58) &  \\ 
  5-5.5 & 4,917 (17.02) & 531 (12.20) & 4,386 (17.87) &  \\ 
  6-6.5 & 9,453 (32.72) & 1,244 (28.57) & 8,209 (33.45) &  \\ 
  7-7.5 & 7,875 (27.25) & 1,368 (31.42) & 6,507 (26.51) &  \\ 
  8-8.5 & 5,577 (19.30) & 1,019 (23.40) & 4,558 (18.57) &  \\ 
  9 & 666 (2.30) & 172 (3.95) & 494 (2.01) &  \\ 
   \hline
Maternal age in years, n (\%) &  &  &  & = 0.0001 \\ 
   \hline
(11,20] & 5,312 (18.38) & 888 (20.40) & 4,424 (18.03) &  \\ 
  (20,30] & 15,010 (51.95) & 2,178 (50.02) & 12,832 (52.29) &  \\ 
  (30,40] & 7,695 (26.63) & 1,181 (27.12) & 6,514 (26.54) &  \\ 
  (40,53] & 878 (3.04) & 107 (2.46) & 771 (3.14) &  \\ 
   \hline
Maternal ethnicity, n (\%) &  &  &  & $<$ 0.0001 \\ 
   \hline
Asian & 2,610 (9.03) & 510 (11.71) & 2,100 (8.56) &  \\ 
  Black & 1,839 (6.36) & 310 (7.12) & 1,529 (6.23) &  \\ 
  Not Reported & 537 (1.86) & 64 (1.47) & 473 (1.93) &  \\ 
  Other & 1,568 (5.43) & 247 (5.67) & 1,321 (5.38) &  \\ 
  White & 22,341 (77.32) & 3,223 (74.02) & 19,118 (77.90) &  \\ 
   \hline
BMI, n (\%) &  &  &  & = 0.5364 \\ 
   \hline
Underweight ($<$18.5) & 2,149 (7.44) & 329 (7.56) & 1,820 (7.42) &  \\ 
  Normal (18.5-24.9) & 14,667 (50.76) & 2,209 (50.73) & 12,458 (50.76) &  \\ 
  Overweight (25.0-29.9)  & 7,367 (25.50) & 1,080 (24.80) & 6,287 (25.62) &  \\ 
  Obese ($>$30.0) & 4,712 (16.31) & 736 (16.90) & 3,976 (16.20) &  \\ 
   \hline
Previous abortions, n (\%) &  &  &  & $<$ 0.0001 \\ 
   \hline
0 & 18,354 (63.52) & 2,949 (67.73) & 15,405 (62.77) &  \\ 
  1-2 & 9,876 (34.18) & 1,301 (29.88) & 8,575 (34.94) &  \\ 
  3+ & 665 (2.30) & 104 (2.39) & 561 (2.29) &  \\ 
   \hline
Previous births, n (\%) &  &  &  & $<$ 0.0001 \\ 
   \hline
0 & 18,354 (63.52) & 2,949 (67.73) & 15,405 (62.77) &  \\ 
  1-2 & 5,070 (17.55) & 609 (13.99) & 4,461 (18.18) &  \\ 
  3+ & 5,471 (18.93) & 796 (18.28) & 4,675 (19.05) &  \\ 
   \hline
Previous Cesarean sections, n (\%) &  &  &  & $<$ 0.0001 \\ 
   \hline
0 & 18,354 (63.52) & 2,949 (67.73) & 15,405 (62.77) &  \\ 
  1-2 & 1,346 (4.66) & 169 (3.88) & 1,177 (4.80) &  \\ 
  3+ & 9,195 (31.82) & 1,236 (28.39) & 7,959 (32.43) &  \\ 
   \hline
Previous miscarriages: &  &  &  & $<$ 0.0001 \\ 
   \hline
0 & 18,354 (63.52) & 2,949 (67.73) & 15,405 (62.77) &  \\ 
  1-2 & 2,046 (7.08) & 264 (6.06) & 1,782 (7.26) &  \\ 
  3+ & 8,495 (29.40) & 1,141 (26.21) & 7,354 (29.97) &  \\ 
   \hline
\end{tabular}
}
\caption{Descriptive characteristics of patients choosing simultaneous or interval administration of mifepristone and misoprostol for early medical abortion.  Patients obtained early medical abortion from May 1, 2015 to April 30, 2016 at British Pregnancy Advisory Service clinics. P-values test for differences in distribution of patient characteristics between the simultaneous and interval groups using chi-squared tests.}
\label{table1}
\end{table}

\begin{centering}\section{Targeted Smooth Bayesian Causal Forests} \label{sec-tsBCF} \end{centering}

\methN{} (\methA{}) offers a promising technique for estimating relative effectiveness of simultaneous versus interval administration of mifepristone and misoprostol over gestation.  
We introduce notation, present the tsBCF model, and discuss assumptions.  
We then review relevant literature.

Let $y$ denote a binary response, and $z$ a binary treatment indicator.  
Let $x$ denote a $p$-length vector of covariates, and $t$ a scalar target covariate over which we induce smoothness.  
Consider a sample of independent observations $\left(y_i, t_i, z_i, x_i \right)$ for $i \in \left\{1,\hdots,n \right\}$.  
Let $\hat{\pi}(x_i)$ be estimates of the propensity score $P(z_i=1 \mid x_i)$.  
In our application, $y_i$ indicates success of the early medical abortion.  
Let $x_i$ represent covariates for patient $i$: maternal age (years), Body Mass Index (kg/m$^2$), maternal ethnicity (Asian, Black, Other, Not Reported, White), and numbers of previous abortions, births, Cesarean sections, and miscarriages.  
Let $t_i$ represent the gestational age at time of early medical abortion, in discrete half-weeks ranging from 4.5 (32 days) to 9 (63 days); $z_i$ indicates the patient's selection of simultaneous versus interval administration of mifepristone and misoprostol.

Our general framework is
\begin{align} \label{eqn:framework}
	P(y_i=1 \mid t_i, x_i, z_i) &= r(t_i, x_i, z_i) \\
	&= \Phi\left(f(t_i, x_i, z_i) \right) \nonumber
\end{align}
where $r$ is some regression function. Our formulation is motivated by two key scientific concerns.  
First, we want to induce smoothness over $t$ while remaining agnostic to smoothness over $x$ and $z$.  
Second, we want to capture three-way interactions between $t$, $x$, and $z$, without specification of their parametric form.  
The probit link function is appropriate for our causal estimand, described below (\ref{eqn:relrisk}).  
Inference is accomplished using latent variable representation and data augmentation \citep{albert:chib:1993}.
\begin{align} \label{eqn:latent-rep}
	y_i &= \begin{cases}
		1 \quad \text{ if } \tilde{y}_i > 0 \\
		0 \quad \text{ if } \tilde{y}_i < 0 \\
	\end{cases}\\
	\tilde{y}_i &= f(t_i, x_i, z_i) + \epsilon_i \nonumber \\
	\quad \epsilon_i &\sim N(0,\sigma^2). \nonumber
\end{align}

Our causal estimand is the conditional average relative effectiveness, or relative risk, of simultaneous versus interval administration. 
\begin{align} \label{eqn:relrisk}
	RR(t,x) =  \frac{ \Phi\left[f(t, x, z=1) \right)}{ \Phi\left(f(t, x, z=0) \right]}
\end{align}

We formulate assumptions using the counterfactual outcomes framework of \citet{imbens2015}, where $y_i(0)$ and $y_i(1)$ denote potential outcomes under interval and simultaneous protocols respectively.  
Observations correspond to realized treatments such that $y_i = z_i y_i(1) + \left(1-z_i \right) y_i(0)$. 
We make several assumptions: non-interference, no unmeasured confounders, and sufficient overlap to estimate treatment effects everywhere in covariate space.  More formally,
~\\
\begin{assump}
We make the stable unit treatment value assumption (SUTVA), which excludes interference between units and multiple versions of treatment \citep{imbens2015}.
\end{assump}
\begin{assump}
	We assume strong ignorability, which specifies that
\begin{gather*}
	\bigl(y_i\left(0 \right), y_i\left(1 \right) \bigr) \indep \bigl( z_i \mid \left(t_i,x_i \right) \bigr) \\ 
	\text{ and } \\
	0 < P\left(z_i=1 \mid t_i,x_i\right) < 1
\end{gather*}
hold for all $i \in 1,\hdots,n$ observations.
\end{assump}

We represent $f(t,x,z)$ as
\begin{align} \label{eqn:tsbcf-model}
	f(t_i,x_i,z_i) &= \mu\left(t_i, x_i, \hat{\pi}_i \right) + z_i \tau \left(t_i, x_i \right)
\end{align}
where $\mu\left(t_i, x_i, \hat{\pi}_i \right)$ and $\tau \left(t_i, x_i \right)$ have independent tsBART priors \citep{starling2019}. For simplicity, we restrict to the mean-zero additive error setting so that
\begin{align*}
	E(\tilde{y}_i \mid t_i, x_i,  z_i) = f(t_i, x_i, z_i).
\end{align*}
Under SUTVA and strong ignorability, we can express the relative risk (\ref{eqn:relrisk}) as
\begin{align}
	RR(t,x) =  \frac{ \Phi\left[ \mu\left(t_i, x_i, \hat{\pi}_i \right) + \tau \left(t_i, x_i \right)  \right]}
					{  \Phi\left[ \mu\left(t_i, x_i, \hat{\pi}_i \right)  \right]}.
\end{align}

Leveraging the tsBART prior induces smoothness in the target covariate; tsBART replaces scalar terminal node parameters with smooth functions over this covariate and assigns Gaussian process priors over the function.  
To model $\mu$, we use a tsBART prior with 200 trees, depth penalty $\beta=2$, splitting probability $\eta=0.95$, and default smoothing parameter $\kappa_\mu=1$, with a half-Cauchy prior on the scale of leaf parameters \citep{gelman2006}.  
For modeling $\tau$, we prefer stronger regularization to reflect our belief that effect of the treatment is generally simpler; we use 50 trees, $\beta=3$, $\eta=0.25$, $\kappa_\tau=1$, and a half-Normal prior on the scale of the tree leaves as in \citet{hahn2020}.  
\citet{starling2019} offer approaches for tuning the tsBART smoothing parameter, and similar tuning may be undertaken here in conjunction with prior knowledge about the likely form of heterogeneity. 
The model is fit via Bayesian backfitting (Appendix \ref{appendix-a2}).

Our work is motivated by Bayesian Causal Forests, or BCF \citep{hahn2020}.  
BCF estimates heterogeneous treatment effects using BART priors, allowing for careful regularization of the model.  
While BCF has desirable properties, there are two fundamental differences between BCF and our method.  
First, BCF lacks a mechanism to induce a priori knowledge of smoothness.  
Second, BCF is formulated for a continuous response; our outcome is binary, and our causal estimand is a non-linear transformation.  
This raises an issue not addressed in the BCF literature, which we call \emph{structural heterogeneity}.  
Heterogeneity in relative risk arises due to heterogeneity in baseline risk, which we term structural, and heterogeneous responses to treatment.
In other words, shrinkage of $\tau$ on the latent scale does not imply shrinkage towards homogeneous relative risks.
Care must be taken in setting sensible prior scales $s_\mu$ and $s_\tau$.  
We propose setting $s_\mu$ based on prior elicitation of a plausible range of baseline risk, and $s_\tau$ using estimation of baseline risk and relative risk from a small amount of held-out data.  
The degree to which structural heterogeneity appears in the prior depends on the average baseline risk, treatment effect size, and variability in $\mu_i$ (Appendix \ref{appendix-a3}, Figure \ref{fig:prob-ill}).

Different covariate vectors $x_i$ may be used in estimation of $\mu$ and $\tau$.  
We specifically include propensity score estimates when fitting $\mu$; we refer readers to \citet{rosenbaum1983, hahn2020} for a discussion in support of including both covariates and propensity score.  
Briefly, propensity score inclusion is an effective dimension-reduction technique yielding a prior that flexibly adapts to complex patterns of confounding, and inclusion of control covariates is necessary when the response does not depend on covariates strictly through the propensity score.  
We estimate propensity scores using BART.

We refer interested readers to \citet{chipman2010} for a detailed review of BART.  
BART has been successful in a variety of contexts including prediction and classification \citep{chipman2010, murray2017log, linero2018, hernandez2018}, survival analysis \citep{sparapani2016, starling2019}, and causal inference \citep{hill2011, hahn2020, logan2017, sivaganesan2017}.   
\citet{linero2017} propose smoothing a regression tree ensemble by randomizing the decision rules at internal nodes of the tree. 
This model induces smoothness over all covariates by effectively replacing the step function induced by the binary trees with sigmoids, instead of smoothing over one targeted covariate.  
Probit versions of BART and tsBART are defined in \citep{chipman2010} and \citep{starling2019}; we draw on these formulations in specifying the model for our binary response.

\begin{centering}\section{Early Medical Abortion Modeling} \label{sec-results} \end{centering}

We now focus on our scientific problem, estimating relative effectiveness of simultaneous versus interval administration of mifepristone and misoprostol across gestation.  
We apply \methA{} to the BPAS data described in Section \ref{sec-ema}.  
Our goal is to model relative effectiveness of the simultaneous versus interval protocols across gestational age, and identify whether there are subgroups of patients for whom the drop in efficacy under simultaneous is markedly wider at later gestations (7--9 weeks). 
We use ``relative effectiveness'' to refer to relative risk (\ref{eqn:relrisk}) when interpreting results, as discussing risk of successful procedure is less clinically intuitive; a value of 0.95 is interpreted as the simultaneous protocol being 95\% as effective as the interval protocol.

For each MCMC iteration (indexed by $b$) of the \methA{} backfitting algorithm, we draw posterior relative risk $RR^{(b)}_{i}$ for patient $i$.  
Averaging across patients observed at each gestational age yields draws of posterior relative risk draws for that gestation.  
Averaging across subgroups of patients at a given gestation yields posterior relative risk draws for that subgroup--gestation combination.  
We summarize results with posterior means and 95\% credible intervals.  
Formally, for MCMC draws $b \in \left\{1,\hdots,B \right\}$, $RR^{(b)}_{i}$ is the $b^{th}$ draw for relative risk for individual $i$ who is observed at gestational age $t_i$.  
We obtain posterior draws of estimated relative effectiveness at gestational age $t$ as
\begin{align} \label{eqn:formal-rr}
	\widehat{RR}^{(b)}_{t} = \sum_{i \mid t_i \in t} \hat{RR}^{(b)}_{i}.
\end{align}
The posterior mean and credible interval for $RR_t$ are calculated using draws $\left\{\widehat{RR}^{(1)}_{t}, \hdots, \widehat{RR}^{(B)}_{t} \right\}$.  Conditioning on $x$ and $t$ in the summation (\ref{eqn:formal-rr}) gives gestation-specific posterior mean relative risk for that subgroup.

\begin{figure}[ht]
		\centering
			\includegraphics[width=0.9\textwidth]{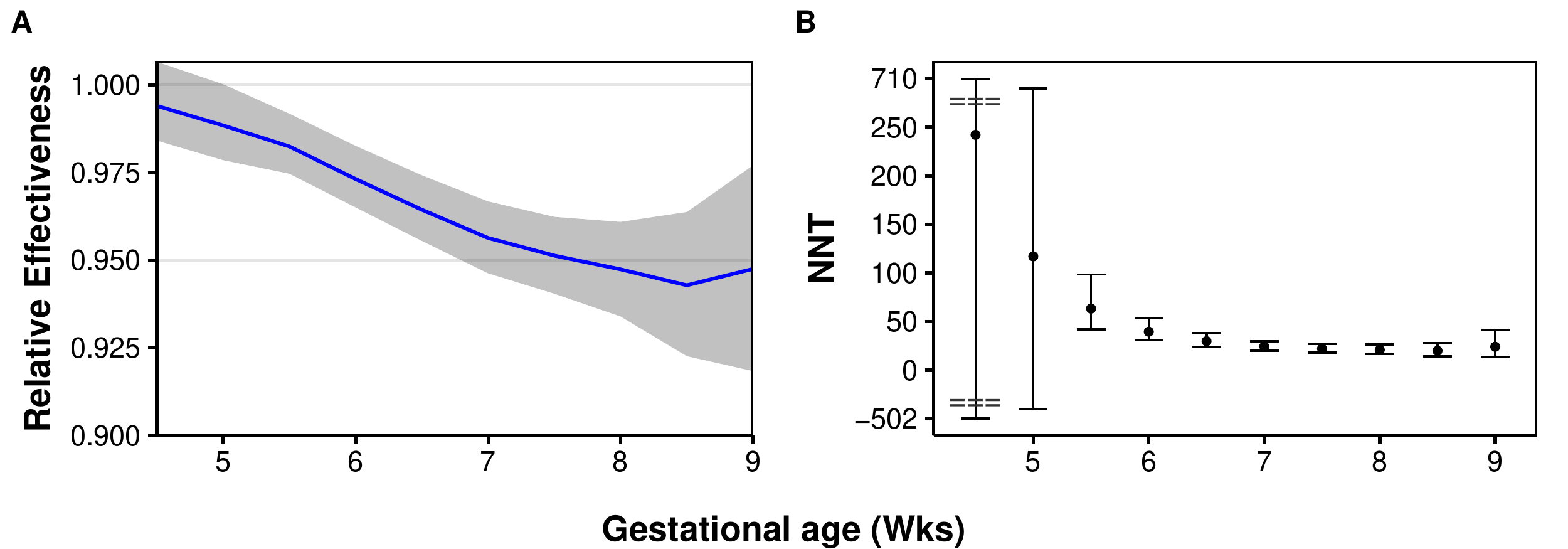}
		\caption{Average relative effectiveness and number needed to treat for the cohort.  (A)  Posterior mean relative effectiveness (solid line) and 95\% credible interval (shaded) for simultaneous versus interval protocols, averaged over all patients, across gestational age. (B) Posterior mean number needed to treat (NNT), giving the average number of patients selecting the simultaneous protocol before one additional failed EMA is observed compared to failures under the interval regimen.  While there is a slight decrease in relative effectiveness as gestation advances, the average relative effectiveness remains high over the course of gestation.}
		\label{fig:ema1}
	\end{figure}

\textbf{Results for cohort.} 
While we find a slight decrease in relative effectiveness at later gestational ages, average relative effectiveness remains high over the course of gestation.  
Average relative effectiveness ranges from 0.994 at 4.5 weeks, to 0.972 at 6 weeks, to 0.955 at 7 weeks, to 0.948 at 9 weeks (Figure \ref{fig:ema1}, Panel A).  
The slight kick-up at week nine is plausibly due to sampling variation; there do not appear to be significant differences in covariates between the patients at eight versus nine weeks (for example, p=.507 for maternal age, and p=.201 for BMI).  
The wider credible interval at week 9 is due to smaller sample size; patients obtaining early medical abortion at 9 weeks account for only 2.3\% of cases (Table \ref{table1}). 
 
We transform these results to number needed to treat (NNT), representing the number of patients who would need to select the simultaneous protocol  before one additional failed procedure is observed, compared to failures under the interval protocol (Figure \ref{fig:ema1}, Panel B).   
The NNT intervals at 4.5 and weeks are large and include zero; the posterior relative risk is close to 1 here, with credible interval bounds falling on each side of one, indicating that simultaneous and interval protocols are nearly equally effective this early in gestation.

We also investigate the degree to which confounding is likely present due to targeted selection \citep{hahn2018}.  We fit a logistic regression model to our binary outcome using estimated propensity scores as the covariate.  The pseudo-$R^2$ \citep{mcfadden1974} is 0.004, indicating lack of evidence for targeted selection.  Finally, we compare variability in posterior relative risk under heterogeneous versus homogeneous $\tau_i$, and find that there is an average of 2.06 times more variability in posterior relative risk under heterogeneous versus homogeneous $\tau_i$, giving us confidence that the following clinical findings for subgroups and individual patients are not just an artifact of structural heterogeneity (Appendix \ref{appendix-a3}).

\begin{figure}[ht]
		\centering
			\includegraphics[width=0.7\textwidth]{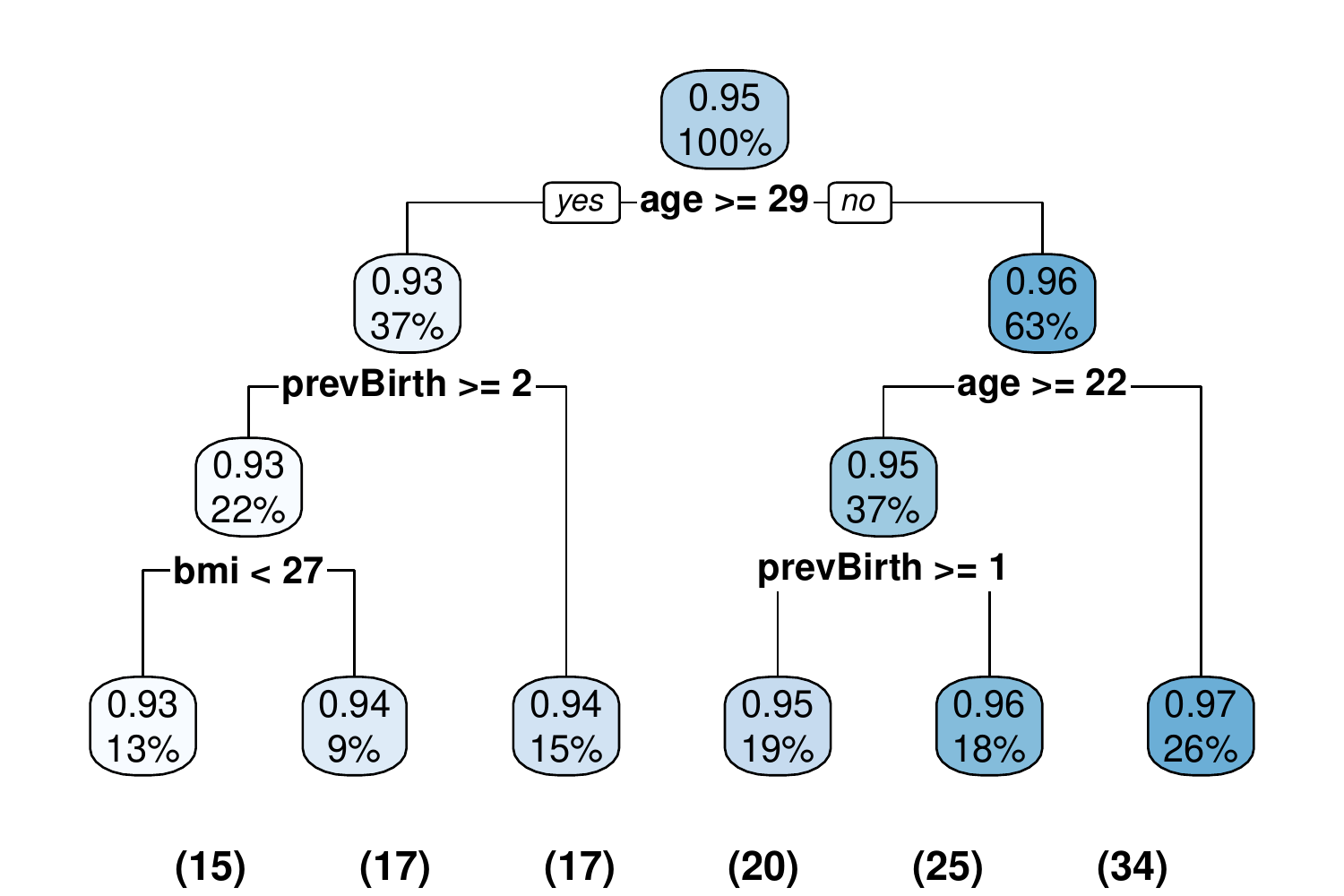}
		\caption{CART tree from the ``fit-the-fit'' subgroup analysis for patients observed at 7--9 weeks gestation.  Each node contains the posterior mean relative effectiveness and the percent of observations contained in that split or terminal node.  In parenthesis below each terminal node is the NNT corresponding to that node's mean relative effectiveness.}
		\label{fig:ema2}
	\end{figure}

\textbf{Subgroup analysis.} We now aim to identify subgroups of patients who may have a larger gap in efficacy at later gestational ages (7--9 weeks).
We take a ``fit-the-fit'' approach to subgroup analysis \citep{hahn2020}, where individual relative effectiveness estimates $\widehat{RR}_{i} = \sum_{b=1}^{B} \widehat{RR}^{(b)}_{i}$ for patients observed at 7--9 weeks are used as the response in a CART model, with covariates $x_i$.  The CART tree fit (Figure \ref{fig:ema2}) identifies maternal age as the most important subgroup; patients 29 and older have somewhat lower relative effectiveness compared to their younger counterparts.  Number of previous births also decreases efficacy, though slightly less so in the younger group of patients.  

\begin{figure}[ht]
		\centering
			\includegraphics[width=0.65\textwidth]{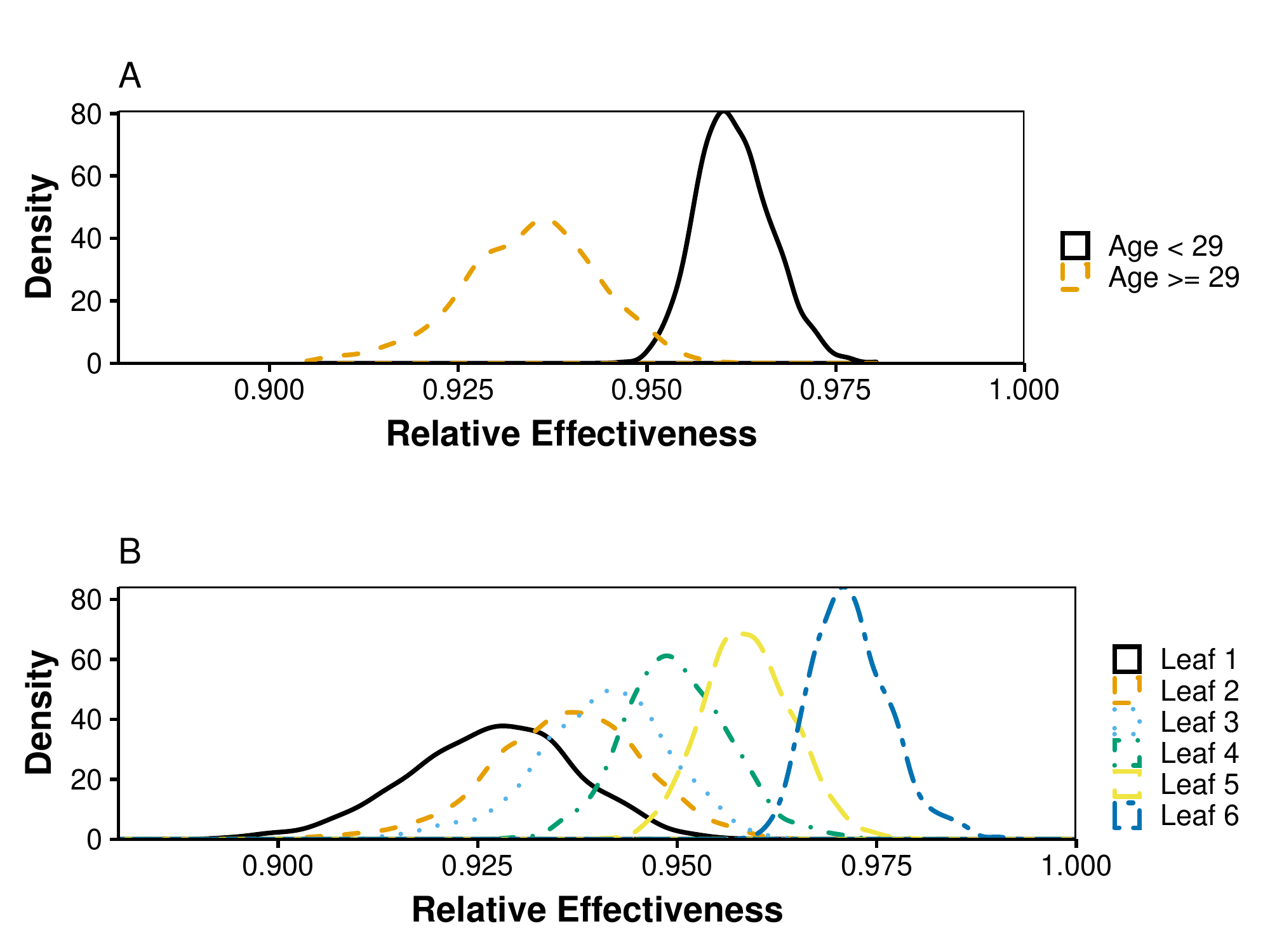}
		\caption{Posterior distributions of relative effectiveness for each CART tree split from Figure \ref{fig:ema2}.  Visualizing tree split posteriors gives insight into significance of differences subgroups of patients.  Age is the most important covariate defining subgroups, with a split at age 29.  Within each age group, previous births decrease relative effectiveness slightly.}
		\label{fig:ema3}
	\end{figure}

The CART fit gives point estimates of relative effectiveness for each node; without uncertainty quantification, it is unclear whether these subgroups are meaningfully different from each other.  
We query the posterior draws for subgroups at each level of the CART tree. Less overlap in posterior densities indicates meaningful splits, and more dispersion indicates greater heterogeneity within a tree split subgroup.  
We plot posteriors for the top split and the terminal nodes (Figure \ref{fig:ema3}).  
The split on age 29 is the most important subgroup, with little overlap in the posteriors.  
Presence of previous births had a slight negative impact on relative effectiveness; this was more pronounced in the older cohort, subtler in patients ages 22--28, and absent in patients under 22. 

To quantify how subgroup differences at later gestations translate to patient impact, we plot the distribution (over MCMC draws) of subgroup average NNT differences between the leftmost and rightmost leaves of the CART tree (Figure \ref{fig:ema2}).  
Differences generally range from 10 to 40 patients, with no mass at zero, indicating that the relative effectiveness gap in these subgroups translates to real difference in number of patients in each subgroup who would need to select the simultaneous protocol at 7--9 weeks before seeing one additional failed procedure.

\begin{figure}[ht]
		\centering
			\includegraphics[width=0.4\textwidth]{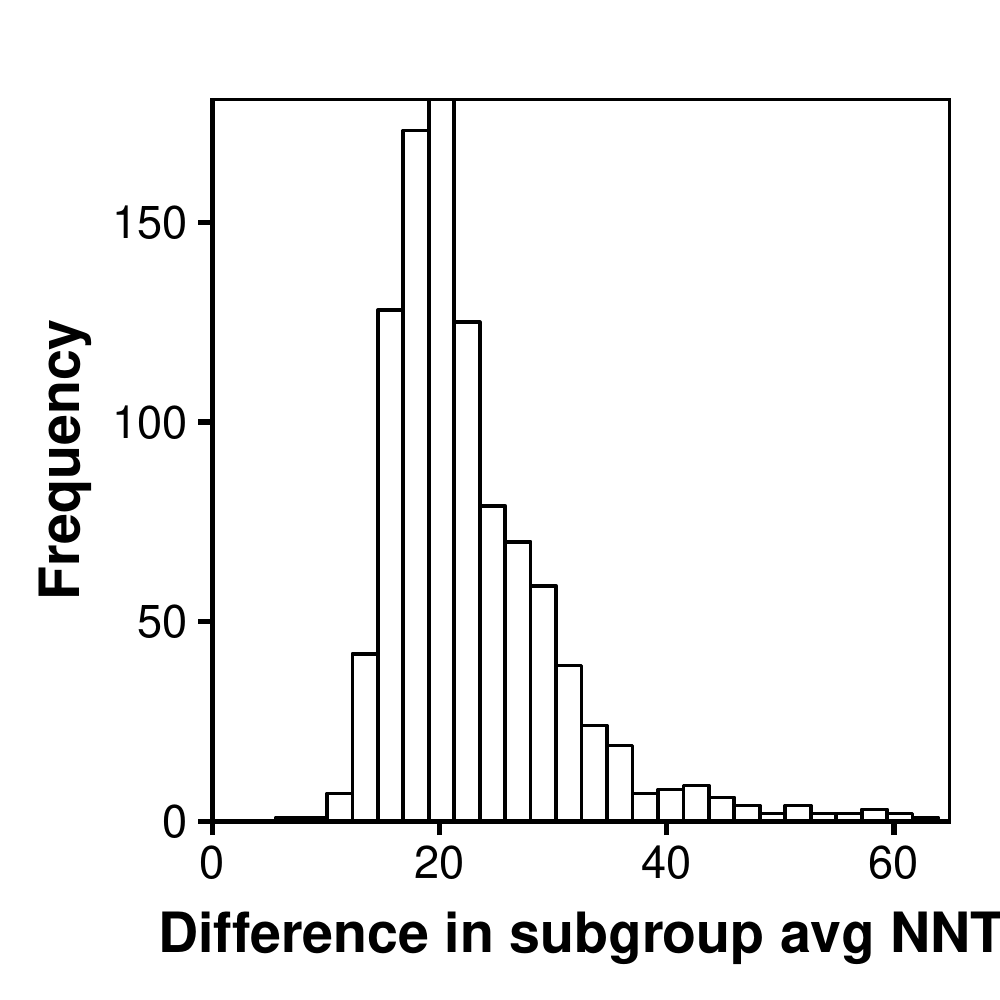}
		\caption{Distribution of differences in subgroup average NNT for the leftmost and rightmost CART tree leaves.  Differences generally range from 10 to 40 patients, indicating that the relative effectiveness gap in these subgroups at later gestational ages translates to real differences in number of patients in each subgroup who would need to select the simultaneous protocol before seeing one additional failed procedure.}
		\label{fig:ema4}
	\end{figure}

\begin{figure}[ht]
		\centering
			\includegraphics[width=0.7\textwidth]{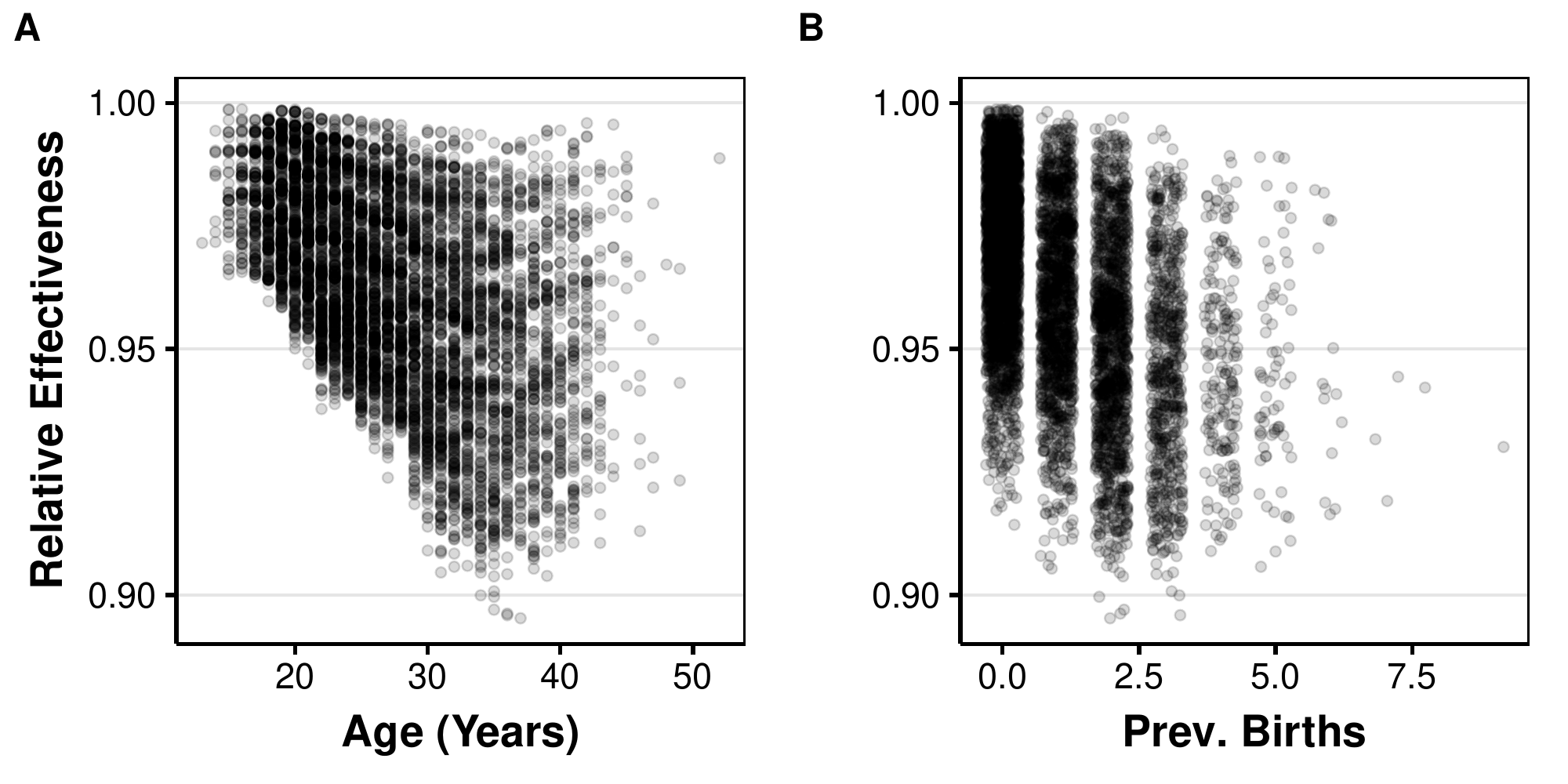}
		\caption{Posterior mean relative effectiveness for each patient, by (A) age and (B) number of previous births.  Increase in both covariates correspond with decreased relative effectiveness; this trend is stronger in age than previous births, consistent with their respective positions in the CART tree.}
		\label{fig:ema5}
	\end{figure}

\textbf{Individual relative effectiveness estimates.} 
Inspecting individual relative effectiveness estimates ensures that there are not smaller subgroups of patients undetected by our analysis, where relative effectiveness is markedly lower.
Figure \ref{fig:ema5} plots posterior mean relative effectiveness for each patient, by age (Panel A) and number of previous births (Panel B), showing the full range of individual relative effectiveness estimates.  
Trends in age and previous births are consistent with our subgroup analysis (Figures \ref{fig:ema2}, \ref{fig:ema3}).
Increase in both covariates correspond with decreased relative effectiveness; this trend is stronger in age than previous births, consistent with their respective positions in the CART tree. 
Table 2 provides detail on cohort characteristics by posterior mean relative effectiveness below 0.90, from 0.90--0.95, and above 0.95. 

\begin{figure}[ht]
		\centering
			\includegraphics[width=0.9\textwidth]{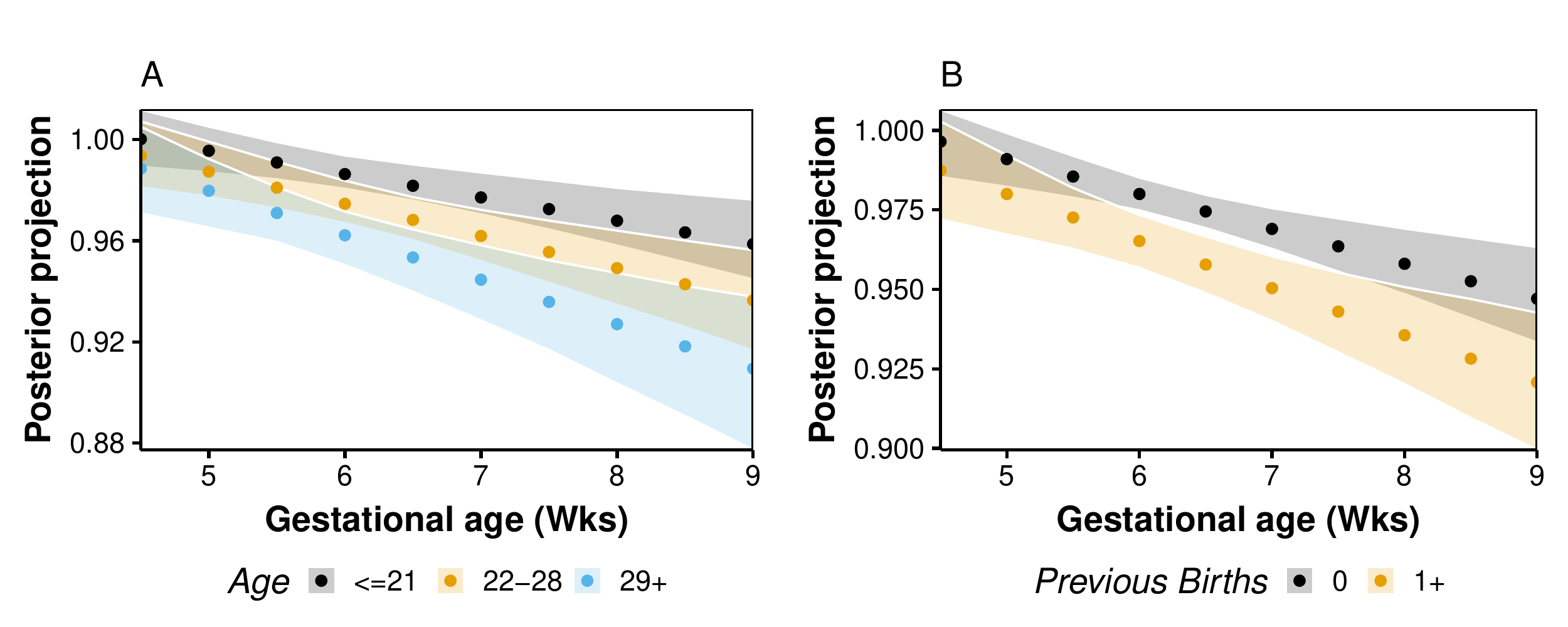}
		\caption{Posterior projection plots for age and previous births.  (A) Posterior projection of relative effectiveness by age group over gestation. 
		(B) Posterior projection of previous births over gestation.  
		The marginal effect of both covariates is consistent with our subgroup analysis.}
		\label{fig:ema6}
	\end{figure}

Posterior projection plots give interpretable lower-dimension model summaries for age and previous births across gestation \citep{woody2019}.  
We project across gestation by age groups (Figure \ref{fig:ema6}, Panel A) and presence of previous births (Figure \ref{fig:ema6}, Panel B), with shaded credible intervals projected from the posterior.  
These plots support our subgroup analysis findings.

\begin{figure}[ht]
		\centering
			\includegraphics[width=0.4\textwidth]{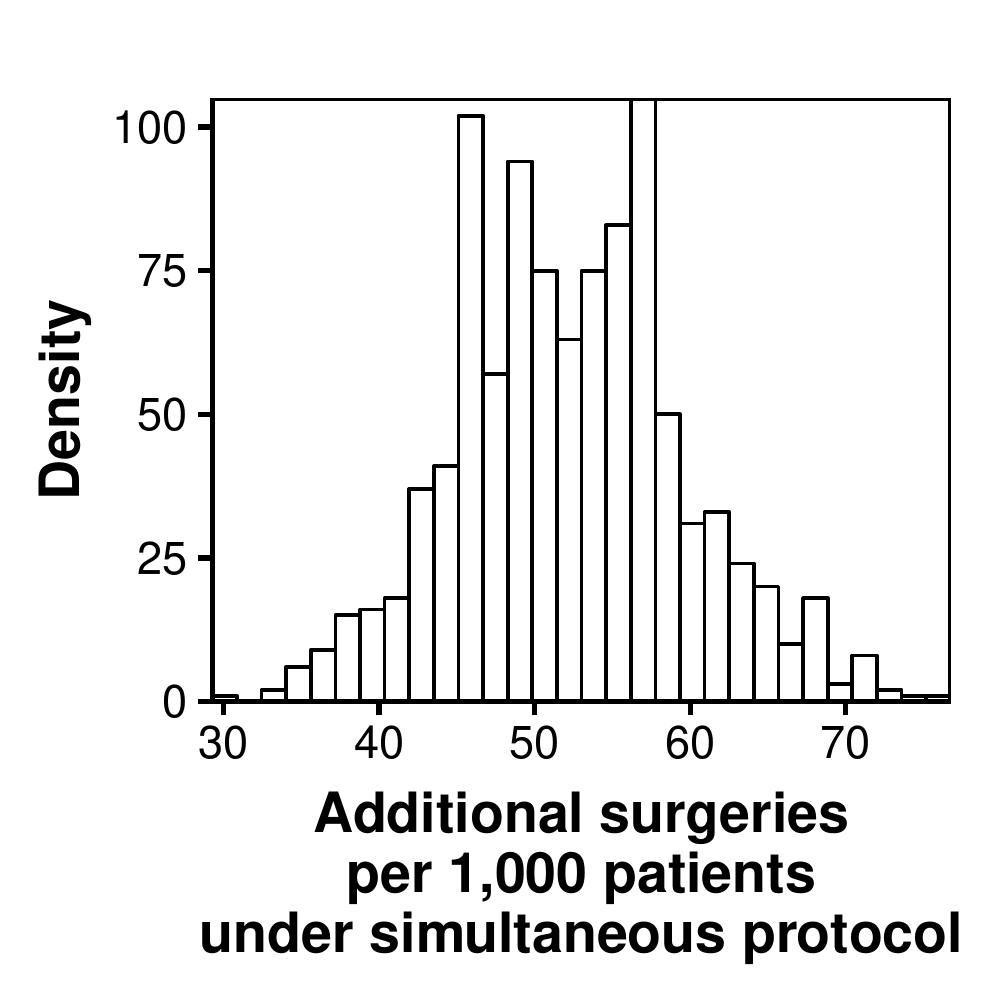}
		\caption{Distribution of differences in observed failures under simultaneous versus expected failure under interval, for patients who experienced a failed procedure under the simultaneous protocol. Clinics choosing to offer the simultaneous protocol can anticipate an additional 40-60 surgeries per 1,000 patients treated.}
		\label{fig:ema7}
	\end{figure}

\textbf{Clinic resource planning.} 
In addition to counseling patients, clinics must plan staff and resources appropriately when offering the simultaneous protocol.
The treatment effect on the treated gives insight here; for patients who experienced a failed procedure under the simultaneous protocol, we plot the distribution of MCMC draws of the differences in the observed number of failures compared to the expected failures had those patients selected the interval protocol (Figure \ref{fig:ema7}).
We report these differences on the order of expected additional surgeries per thousand patients, giving clinics a sense of the likely volume of procedures.
We find that a clinic can anticipate approximately 40--60 additional surgeries per thousand patients per 1,000 patients treated when offering the simultaneous protocol.

\begin{figure}[ht]
		\centering
			\includegraphics[width=0.7\textwidth]{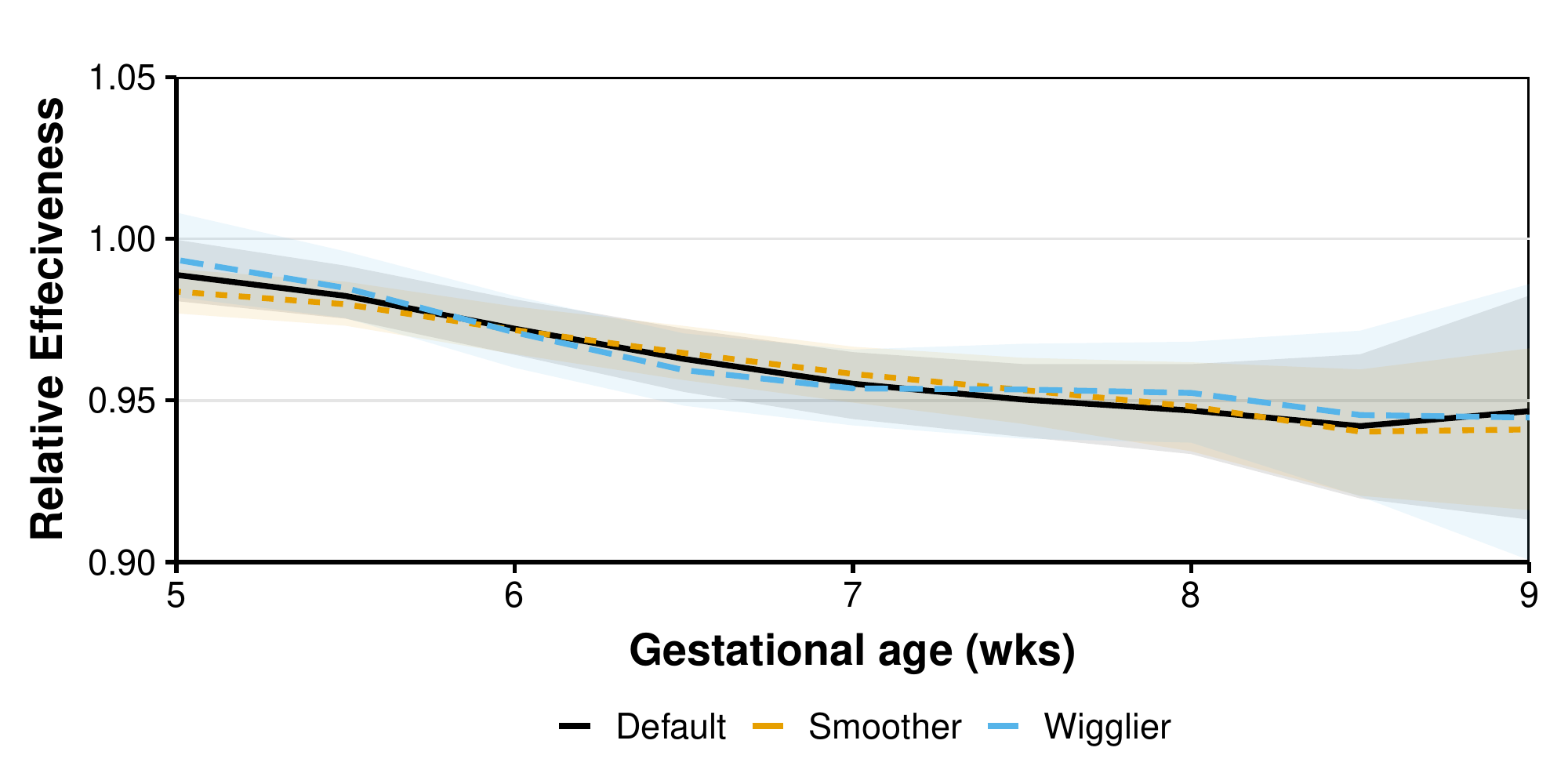}
		\caption{Posterior mean relative effectiveness for three settings of the tsBCF model's $\kappa_\tau$ smoothness parameter.  
		The solid line gives the posterior mean relative effectiveness over gestation from our analysis (Default), with shaded 95\% credible interval.
		Dashed lines give posterior means for $\kappa_\tau=1/3$ (Smoother) and $\kappa_\tau=3$ (Wigglier). 
		We do not see clinically meaningful differences in the three fits across gestation, indicating robustness to choice of smoothness parameter.}
		\label{fig:ema8}
	\end{figure}

\textbf{Sensitivity to smoothing parameters.} 
The early medical abortion analysis used our suggested default smoothness parameter settings ($\kappa_\mu=1$ and $\kappa_\tau=1$). 
Here, we perform a sensitivity analysis for robustness of our analysis to smoothness parameter choice.
We let $\kappa_\mu=1$ and fit the tsBCF model to the early medical abortion dataset three times, for $\kappa_\tau  \in \left\{\frac{1}{3}, 1, 3 \right\}$.
These choices represent a three-times change in magnitude in each direction from the default, corresponding to varying the length-scale of the treatment trees' covariance from one to three to nine.
While there are small differences in the overall estimated relative effectiveness (Figure \ref{fig:ema8}), we do not see clinically meaningful variation across smoothness parameter settings, indicating that our analysis is robust to choice of smoothing parameter. 
The small differences in week 9 lend support to our intuition that the slight kick-up is due to small sample size.\\

\begin{table}[ht]
\centering
\scalebox{0.65}{
\begin{tabular}{lrrrrr}
 Characteristic & (N = 28,895) & RE $>$0.95 (N = 21,304) & RE 0.90-0.95 (N = 7,565) & RE $<$.90 (N = 26) & P-value \\ 
  \hline
Gestational age in weeks, n (\%) &  &  &  &  & $<$ 0.0001 \\ 
   \hline
4.5 & 407 (1.41) & 407 (1.91) & 0 (0.00) & 0 (0.00) &  \\ 
  5-5.5 & 4,917 (17.02) & 4,917 (23.08) & 0 (0.00) & 0 (0.00) &  \\ 
  6-6.5 & 9,453 (32.72) & 8,430 (39.57) & 1,023 (13.52) & 0 (0.00) &  \\ 
  7-7.5 & 7,875 (27.25) & 4,583 (21.51) & 3,291 (43.50) & 1 (3.85) &  \\ 
  8-8.5 & 5,577 (19.30) & 2,605 (12.23) & 2,947 (38.96) & 25 (96.15) &  \\ 
  9 & 666 (2.30) & 362 (1.70) & 304 (4.02) & 0 (0.00) &  \\ 
   \hline
Maternal age in years, n (\%) &  &  &  &  & $<$ 0.0001 \\ 
   \hline
(11,20] & 5,312 (18.38) & 5,312 (24.93) & 0 (0.00) & 0 (0.00) &  \\ 
  (20,30] & 15,010 (51.95) & 12,011 (56.38) & 2,999 (39.64) & 0 (0.00) &  \\ 
  (30,40] & 7,695 (26.63) & 3,485 (16.36) & 4,185 (55.32) & 25 (96.15) &  \\ 
  (40,53] & 878 (3.04) & 496 (2.33) & 381 (5.04) & 1 (3.85) &  \\ 
   \hline
Maternal ethnicity, n (\%) &  &  &  &  & $<$ 0.0001 \\ 
   \hline
Asian & 2,610 (9.03) & 1,941 (9.11) & 669 (8.84) & 0 (0.00) &  \\ 
  Black & 1,839 (6.36) & 1,700 (7.98) & 139 (1.84) & 0 (0.00) &  \\ 
  Not Reported & 537 (1.86) & 400 (1.88) & 137 (1.81) & 0 (0.00) &  \\ 
  Other & 1,568 (5.43) & 1,165 (5.47) & 402 (5.31) & 1 (3.85) &  \\ 
  White & 22,341 (77.32) & 16,098 (75.56) & 6,218 (82.19) & 25 (96.15) &  \\ 
   \hline
BMI category (kg/m$^2$), n (\%) &  &  &  &  & $<$ 0.0001 \\ 
   \hline
Underweight ($<$18.5) & 2,149 (7.44) & 1,716 (8.05) & 433 (5.72) & 0 (0.00) &  \\ 
  Normal (18.5-24.9) & 14,667 (50.76) & 10,878 (51.06) & 3,769 (49.82) & 20 (76.92) &  \\ 
  Overweight (25.0-29.9)  & 7,367 (25.50) & 5,179 (24.31) & 2,182 (28.84) & 6 (23.08) &  \\ 
  Obese ($>$30.0) & 4,712 (16.31) & 3,531 (16.57) & 1,181 (15.61) & 0 (0.00) &  \\ 
   \hline
Previous abortions, n (\%) &  &  &  &  & $<$ 0.0001 \\ 
   \hline
0 & 18,354 (63.52) & 14,373 (67.47) & 3,980 (52.61) & 1 (3.85) &  \\ 
  1-2 & 9,876 (34.18) & 6,546 (30.73) & 3,306 (43.70) & 24 (92.31) &  \\ 
  3+ & 665 (2.30) & 385 (1.81) & 279 (3.69) & 1 (3.85) &  \\ 
   \hline
Previous births, n (\%) &  &  &  &  & $<$ 0.0001 \\ 
   \hline
0 & 18,354 (63.52) & 14,373 (67.47) & 3,980 (52.61) & 1 (3.85) &  \\ 
  1-2 & 5,070 (17.55) & 3,071 (14.42) & 1,989 (26.29) & 10 (38.46) &  \\ 
  3+ & 5,471 (18.93) & 3,860 (18.12) & 1,596 (21.10) & 15 (57.69) &  \\ 
   \hline
Previous Cesarean cections, n (\%) &  &  &  &  & $<$ 0.0001 \\ 
   \hline
0 & 18,354 (63.52) & 14,373 (67.47) & 3,980 (52.61) & 1 (3.85) &  \\ 
  1-2 & 1,346 (4.66) & 801 (3.76) & 545 (7.20) & 0 (0.00) &  \\ 
  3+ & 9,195 (31.82) & 6,130 (28.77) & 3,040 (40.19) & 25 (96.15) &  \\ 
   \hline
Previous miscarriages: &  &  &  &  & $<$ 0.0001 \\ 
   \hline
0 & 18,354 (63.52) & 14,373 (67.47) & 3,980 (52.61) & 1 (3.85) &  \\ 
  1-2 & 2,046 (7.08) & 1,082 (5.08) & 948 (12.53) & 16 (61.54) &  \\ 
  3+ & 8,495 (29.40) & 5,849 (27.45) & 2,637 (34.86) & 9 (34.62) &  \\ 
   \hline
\end{tabular}
}
\caption{
Descriptive characteristics by ranges of posterior mean individual relative effectiveness.
P-values assess differences in distribution of patient characteristics between the three ranges of relative effectiveness.
Only 26 patients (0.09\%) have estimated relative effectiveness less than 0.90; all are over 30, and all but one have at least one previous birth.
Patients in the 0.90--0.95 category are also predominantly over 30.
Additionally, 21,304 patients (74\%) have estimated relative effectiveness greater than 0.95. 
}
	\label{table2}
\end{table}

\begin{centering}\section{Simulations} \label{sec-sims} \end{centering}

We compare tsBCF to several existing models in a benchmarking study designed to simulate five clinically plausible scenarios with a binary response. 
We generate latent-scale prognostic effects, treatment effects, and random noise for each scenario, and assess how well the models recover relative risk.
Our goal is to verify that tsBCF successfully estimates heterogeneous relative risks with reasonable uncertainty quantification and nominal coverage, while inducing smoothness over the target covariate. We compare the following models.
\begin{itemize}
	\item \textbf{tsBCF1: } The \methA{} method with default smoothing parameters $\kappa_\mu=1$ and $\kappa_\tau=1$.
	\item \textbf{tsBCF2: } The \methA{} method with smoothing parameters $\kappa_\mu=1$ and $\kappa_\tau=3$.	
	\item \textbf{BCF: } The Bayesian Causal Forest model described in \citet{hahn2020}.  We expect this model to perform well but lack smoothness (Figure 7).
	\item \textbf{BART: } Ordinary BART used to model the response surface in the causal inference setting \citep{hill2011} with estimated propensity scores included as a covariate \citep{rosenbaum1983}.
\end{itemize}

Simulated data is generated as follows.  For independent observations $i \in \left\{1,.\hdots,n \right\}$, draw a vector of covariates $x_i = \left\{x_{1i}, x_{2i}, x_{3i}, x_{4i}, x_{5i} \right\} \stackrel{iid}{\sim} N(0,1)$ and draw target covariate $t_i \in \left\{0.1,0.2,.\hdots,1 \right\}$.  We generate observations from
\begin{align}
	P(y_i=1 \mid t_i, x_i, z_i) = \Phi \left(\mu(t_i,x_{1i},x_{2i}) + \tau(t_i,x_{3i})z_i \right)
\end{align}
where $\mu$ is the prognostic function and $\tau$ is the treatment effect function.  The prognostic function is the same for all five scenarios:
\begin{align}
	\label{eqn:sim-mu-function}
	\mu(t_i,x_{1i},x_{2i}) = \frac{1}{4} t_i^{1.5} + \frac{x_{1i}}{6} + \frac{x_{2i}}{4}
\end{align}

Each observation is assigned to treatment ($z_i=1$) or control ($z_i=0$) based a binomial draw with propensity score
\begin{align}
	\pi_i = \Phi\left( 
		\rho \cdot \left[\frac{x_{1i}}{6}-\frac{x_{2i}}{4} \right] + 
			\left(1-\rho \right) \left[-1\left(x_{4i}>5 + 1(x_{4i}<5) \right) \right]
	\right)
\end{align}
where $\rho \in \left[0,1 \right]$ controls the degree to which the propensity score is based on a somewhat accurate prediction of the potential outcome, since $\left[\frac{x_{1i}}{6}-\frac{x_{2i}}{4} \right]$ is found in the prognostic effect (\ref{eqn:sim-mu-function}), while $x_{4i}$ and $x_{5i}$ are not used in prognostic or treatment effect generation.   We estimate propensity scores using the \textbf{dbarts} R package \citep{chipman2010}; any accurate prediction is viable \citep{hahn2020}.  We set $\rho=0.25$ to introduce a small amount of targeted selection. 

We vary $\tau(t_i,x_i)$ by scenario, reflecting different latent-scale treatment effects as follows.
\begin{itemize}
	\item Scenario A represents a treatment effect that varies smoothly over the target covariate $t$, with homogeneity in $x$.
	\begin{align*}
		\tau(t_i,x_i) = 0.1 + 0.2 t_i - 0.05\sin(1.5 \pi t_i)
	\end{align*}
	\item Scenario B represents heterogeneous treatment effects that vary smoothly over $t$ with modest differences in subgroups.
	\begin{align*}
		\tau(t_i,x_i) = 0.1 + 0.2\mathbb{I}\left(x_{3i} > -1/2 \right) + 0.15 \mathbb{I}\left(x_{3i} > 1/2 \right) + 0.2 t_i - 0.05 \sin(1.5 \pi t_i)
	\end{align*}
	\item Scenario C represents heterogeneous treatment effects, similar to Scenario B except that the effects of $t$ and $x$ are inseparable.
	\begin{align*}
		\tau(t_i,x_i) = 0.1 + 0.2\mathbb{I}\left(x_{3i} > -1/2 \right) + \left(0.15 + 0.2 t_i \right) \mathbb{I}\left(x_{3i} > 1/2 \right) +
		0.2 t_i - 0.05 \sin(1.5 \pi t_i)
	\end{align*}
	\item Scenario D gives heterogeneous treatment effects with small effects in general, except for a small subgroup with a pronounced effect.
	\begin{align*}
		\tau(t_i,x_i) = 0.05 + 0.05\mathbb{I}\left(x_{3i} > -1/2 \right) + \left(0.15 + 0.2 t_i \right) \mathbb{I}\left(x_{3i} > 1/2 \right) +
		0.2 t_i - 0.05 \sin(1.5 \pi t_i)
	\end{align*}
	\item Scenario E is a constant treatment effect, requiring shrinking to homogeneity in both $x$ and $t$.
	\begin{align*}
		\tau(t_i,x_{3i}) = 0.1
	\end{align*}
\end{itemize}

For 50 replicates of each scenario, we generate a dataset and fit all models for each scenario. 
Table \ref{tbl:sim-scenarios} gives results averaged across replicates for each scenario. 
RMSE is the average root mean squared error for estimating heterogeneous relative risk; nominal coverage is 95\%, and interval length is for posterior credible interval of the relative risk estimates for each unit.  
TsBCF and BCF perform comparably in RMSE, coverage, and interval length across scenarios.  
BART performs nearly as well but over-covers.  
Differences are small compared to standard error size; these results confirm that in introducing smoothness in a target covariate adds interpretability without compromising performance.

\begin{figure}[ht]
		\centering
			\includegraphics[width=0.9\textwidth]{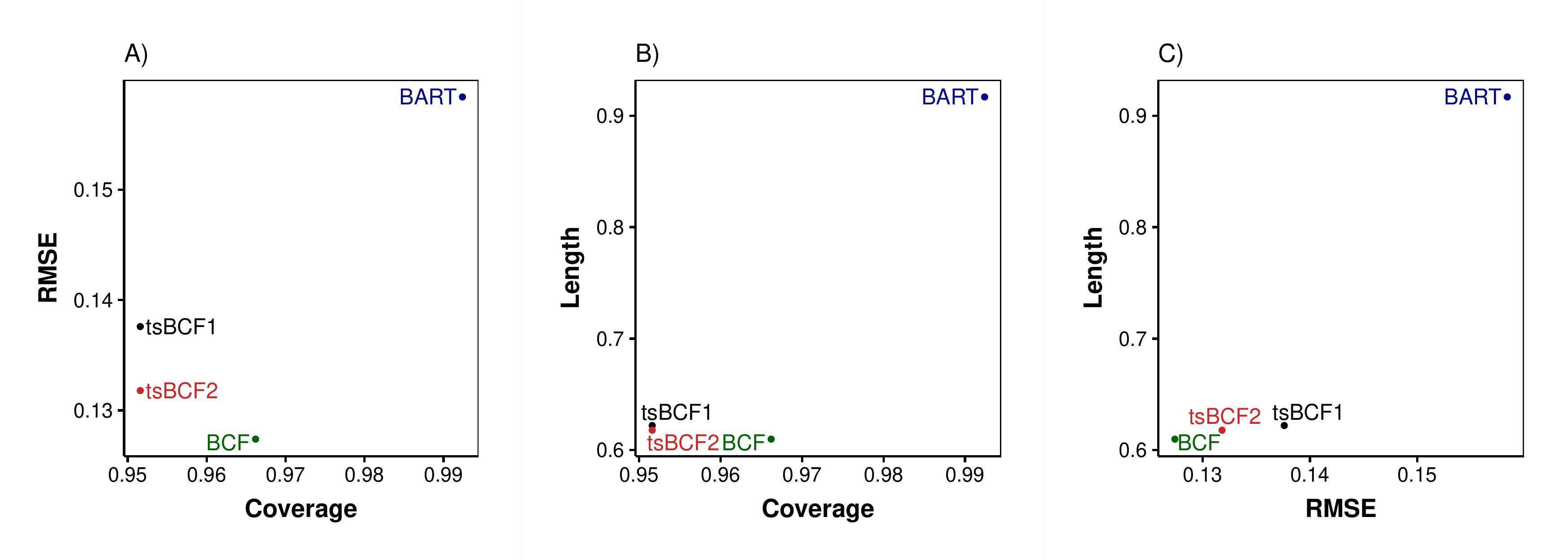}
		\caption{Simulation results for each method.  Panel A gives coverage versus RMSE.  Panel B gives coverage versus interval length.  Panel C gives RMSE versus interval length. The tsBCF models recover the heterogeneous relative risks while maintaining coverage and yielding reasonable measures of uncertainty.}
		\label{fig:sim-fig}
	\end{figure}

Figure \ref{fig:sim-fig} illustrates the relationships of RMSE, coverage, and interval length.  Panel A gives coverage versus RMSE, where the tsBCF methods combine low RMSE with good coverage and the difference between tsBCF and BCF is negligible.  Panel B gives coverage versus interval length, where tsBART and BCF are maintaining nominal coverage and have similar interval lengths.  Panel C gives RMSE versus interval length, where tsBCF and BCF are performing comparably.  Values are averaged over scenarios. Together, these panels demonstrate that tsBCF recovers the heterogeneous relative risks while maintaining coverage and yielding reasonable measures of uncertainty in scenarios when the underlying treatment effect is smooth over the target covariate.

\begin{table}[ht]
\centering
\begin{tabular}{llrrrr}
  \hline
Scenario & Model & RMSE & SD(RMSE) & Coverage & Interval Length \\ 
  \hline
A & tsBCF1 & 0.104 & 0.030 & 0.951 & 0.526 \\ 
   & tsBCF2 & 0.095 & 0.037 & 0.951 & 0.519 \\ 
   & BCF & 0.096 & 0.032 & 0.968 & 0.522 \\ 
   & BART & 0.121 & 0.052 & 0.997 & 0.844 \\ 
      \hline 
  B & tsBCF1 & 0.173 & 0.046 & 0.940 & 0.708 \\ 
   & tsBCF2 & 0.166 & 0.046 & 0.940 & 0.704 \\ 
   & BCF & 0.159 & 0.038 & 0.959 & 0.692 \\ 
   & BART & 0.187 & 0.063 & 0.990 & 0.994 \\ 
      \hline
  C & tsBCF1 & 0.191 & 0.056 & 0.947 & 0.797 \\ 
   & tsBCF2 & 0.185 & 0.048 & 0.947 & 0.772 \\ 
   & BCF & 0.174 & 0.040 & 0.968 & 0.770 \\ 
   & BART & 0.206 & 0.065 & 0.988 & 1.038 \\ 
      \hline 
  D & tsBCF1 & 0.161 & 0.042 & 0.930 & 0.634 \\ 
   & tsBCF2 & 0.151 & 0.040 & 0.930 & 0.634 \\ 
   & BCF & 0.148 & 0.039 & 0.946 & 0.621 \\ 
    & BART & 0.172 & 0.060 & 0.988 & 0.911 \\ 
      \hline
  E & tsBCF1 & 0.059 & 0.040 & 0.990 & 0.445 \\ 
   & tsBCF2 & 0.062 & 0.039 & 0.990 & 0.460 \\ 
   & BCF & 0.060 & 0.042 & 0.990 & 0.444 \\ 
   & BART & 0.106 & 0.052 & 0.999 & 0.797 \\ 
   \hline
\end{tabular}
\caption{Simulation results averaged across replicates for each scenario and model. RMSE is the average root mean squared error for estimating heterogeneous relative risk; nominal coverage is 95\%, and interval length is for posterior credible interval of the relative risk estimates for each unit.  TsBCF and BCF perform comparably in RMSE, coverage, and interval length across scenarios.  BART performs nearly as well but over-covers.}
	\label{tbl:sim-scenarios}
\end{table}

\begin{centering}\section{Discussion} \label{sec-discussion} \end{centering}

Targeted Smooth Bayesian Causal Forests allows for estimation of heterogeneous treatment effects which evolve smoothly over a target covariate.  
This addresses a key statistical issue that arises in virtually all pregnancy-related research: that most outcomes vary smoothly with gestational age. 
Our model is nonparametric, and so avoids potential bias arising from specification of functional forms of the causal estimand.  
TsBCF enjoys similar advantages as BCF, including hyperparameters which are set efficiently using heuristics similar to \citet{hahn2020} and \citet{chipman2010}.  
Like tsBART, tsBCF has easily-tuned hyperparameters to control degree of smoothness, with default settings yielding excellent performance \citep{starling2019}.

Our analysis of the early medical abortion data using the tsBCF model answers key clinical questions which were not addressed satisfactorily in previous research.
We validate previous findings~\citep{lohr2018} that on average, we do not see a marked decrease in efficacy as gestational age increases.
However, we do find a modest decrease in the 7--9 week range (from 0.972 at 6 weeks, to 0.955 at 7 weeks, to 0.948 at 9 weeks).  
We identify a more pronounced drop in efficacy at later gestations for patients age 29 and older, particularly for those who have given birth previously.
While relative effectiveness for these patients is still over 90\% in the 7--9 week range, clinicians may wish to counsel their patients accordingly.

A limitation of our work is the relatively small set of available covariates; it is possible that unobserved confounders exist.  
We do not know what covariates influenced each patient's choice in protocol \citep{lohr2018}.
Patients received counseling on the expected differences in effectiveness and side effects based on a small BPAS pilot study; aside from clinicians' use of a common comparison chart in a printed patient guide, counseling is not standardized.
However, our propensity score estimates reflect the similar safety profiles of both protocols, which share low rates of significant adverse effects, as well as the simultaneous protocol's reduced burden of care.  
Nonetheless, our work presents a clearer and more nuanced picture of the relative efficacy of simultaneous versus interval protocols for administration of mifepristone and misoprostol than previously available.\\

\clearpage
\begin{centering}\section*{REFERENCES}\end{centering}
\begin{small}\
	\bibliographystyle{abbrvnat}
	\bibliography{tsBCF-paper}
\end{small}
		
\newpage
\appendix
\begin{centering}\section{Appendix} \label{sec-appendix} \end{centering}

\subsection{Fitting the \methA{} model using data augmentation.} \label{appendix-a1}
Here we provide details on the model parameterization and prior used for fitting \methA{}.  We rewrite (\ref{eqn:tsbcf-model}) using a redundant multiplicative parameterization \citep{gelman2006}.
\begin{align*}
	f(t_i,x_i,z_i) &=  \alpha_t + \xi \mu(t_i, x_i, \hat{\pi}_i) + 
	\left[b_1 z_i + b_0 \left(1-z_i \right) \right] \tau(t_i, x_i)
\end{align*}
where the treatment effect on the latent probit scale is now $(b_1-b_0)\tau(t_i, x_i)$, and the $\alpha_t$ act as target-specific offsets which we estimate directly from the data.

Priors are as follow, with hyperparameters $\left\{\lambda, s_\mu, s_\tau, l_\mu, l_\tau \right\}$ set by the user. We set $\nu_\mu=1$ to induce the half-Cauchy prior, and $\nu_\tau$ is set to induce the half-Normal prior.
\begin{gather*}
	\xi \sim N(0,1)\\
	b_1 \sim N(.5,.5) \text{ and } b_0 \sim N(-.5,.5) \rightarrow b = (b1-b0) \sim N(0,1) \\
	\mu(t_i, x_i, \hat{\pi}_i) \sim tsBART \text{ with } n_\mu=200 \text{ trees, terminal nodes } m_\mu \text{, and covariance } C_\mu \text{:}\\
	m_\mu \sim Gp(0,C_\mu) \\
	C_\mu = \frac{s^2_\mu}{n_\mu \Delta_\mu} \exp\left[-.5\left(\frac{t-t'}{l_\mu} \right)^2 \right] \\
	\frac{1}{\Delta_\mu} \sim IG\left(\frac{\nu_\mu}{2}, \frac{\nu_\mu}{2} \right) \\
	\tau(t_i, x_i)  \sim tsBART \text{ with } n_\tau=200 \text{ trees, terminal nodes } m_\tau \text{, and covariance } C_\tau \text{:}\\
	m_\tau \sim Gp(0,C_\tau) \\
	C_\tau= \frac{s^2_\tau}{n_\tau \Delta_\tau}  \exp\left[-.5\left(\frac{t-t'}{l_\tau} \right)^2 \right]\\
	\frac{1}{\Delta_\tau} \sim IG\left(\frac{\nu_\tau}{2}, \frac{\nu_\tau}{2} \right)
\end{gather*}

We constrain $\sigma^2=1$ for identifiability. For a continuous response, the prior for $\sigma^2$ follows Chipman et al.'s recommendation for a rough over-estimation of $\hat{\sigma}$.  We choose $\nu=3$ and $q = 0.90$, and estimate $\hat{\sigma}$ by regressing $y$ onto $x$ (including the target variable as a covariate), then choose $\lambda$ s.t. the $q$th quantile of the prior is located at $\hat{\sigma}$, i.e. $P(\sigma \le \hat{\sigma}) = q$. Length-scale parameters $\left(l_\mu, l_\tau \right)$ can be stated in terms of expected wiggliness of the tsBART fits, as described in \citet{starling2019}.

\subsection{Bayesian Backfitting Algorithm} \label{appendix-a2}
We leverage the Bayesian backfitting MCMC algorithms of tsBART and BCF to design a Bayesian backfitting algorithm for \methA{}.  We refer interested readers to \citet{chipman2010} for a full discussion of the original Bayesian backfitting, and \citet{starling2019} and \citet{hahn2020} for tsBART and BCF algorithms respectively.  Briefly, Bayesian backfitting involves an MCMC algorithm where each tree, and its parameters are sampled one at a time given the partial residuals from the other $m-1$ trees.  One iteration of the sampler consists of looping through the trees, sampling each tree $T_j$ via a Metropolis step, and then sampling its associated leaf parameters $M_j$, conditional on $\sigma^2$ and the remaining trees and leaf parameters.  After a pass through all trees, $\xi$, $\Delta_\mu$, $b_1$, $b_0$, $\Delta_\tau$, and $\sigma^2$ are updated in a Gibbs step.

\subsubsection{Updating trees and leaves}
In general, to sample $\left\{T_j,M_j \right\}$ conditioned on the other trees and leaf parameters $\left\{T_{(j)}, M_{(j)} \right\}$, define the partial residual as
\begin{align}
	\label{eqn:resids}
	r_{ij} = y_i - \sum_{k=1, k \neq j}^{m} g(x_i;T_k,M_k) \, .
\end{align}

More specifically in the \methA{} setting, for $\mu(t_i,x_i,\hat{\pi}_i)$:
\begin{align*}
	&\text{The ``data'' is } \left( \frac{y_{i}-\alpha_t-\left[b_1 z_i + b_0 \left(1-z_i \right) \right] \tau(t_i,x_i) }{\xi} \right)\\
	&\text{The variance is } \left(\frac{\sigma^2}{\xi^2} \right) \\
	&\text{The partial residuals are } r_{ij} = \text{data}_i - \sum_{k=1, k \neq j}^{200} g\left(x_k; T_{\mu k}, M_{\mu k } \right).
\end{align*}

Similarly, for $\tau(t_i,x_i)$:
\begin{align*}
	&\text{The ``data'' is } \left(\frac{y_{i}-\alpha_t-\xi \mu(t_i,x_i,\hat{\pi}_i)   }{\left[b_1 z_i + b_0 \left(1-z_i \right) \right]} \right)\\
	&\text{The variance is } \left(\frac{\sigma^2}{\left[b_1 z_i + b_0 \left(1-z_i \right) \right]^2} \right) \\
	&\text{The partial residuals are } r_{ij} = \text{data}_i - \sum_{k=1, k \neq j}^{50} g\left(x_k; T_{\tau k}, M_{\tau k } \right).
\end{align*}


Using $r_j$ as the working response vector, at step $s$ of the MCMC one samples $T_j^{(s)}$ by proposing one of four local changes to $T_j^{(s-1)}$, marginalizing analytically over $M_j$.  The local change is selected randomly from the following candidates: 
\begin{itemize}
	\item \textbf{grow} randomly selects a terminal node and splits it into two child nodes
	\item \textbf{prune} randomly selects an internal node with two children and no grandchildren, and prunes the children, making the selected node a leaf
\end{itemize}

In line with common practice, we implement only \textbf{prune} and \textbf{grow} proposals due to computational cost \citep{pratola2014}. Once the move in tree space is either accepted or rejected, $M_j$ is sampled from its full conditional given  $T_j$ and $\sigma^2$. 

\subsubsection{Full conditionals}
The posterior conditional distributions for the Bayesian backfitting algorithm are as follows.  For simplicity we assume that target covariate values $t$ are on a common discrete grid, though this is not a requirement.

~\\
For updating $\left(\sigma^2 \middle| \hdots \right)$,
\begin{gather*}
	p\left(\sigma^2  \right)\sim \text{IG}\left(\frac{\nu}{2}, \frac{\nu\lambda}{2} \right) \\ 
	%
	p\left(y \middle| \sigma^2 \right) = \prod_{i=1}^{n} \text{N}\left(y_i \middle| \hat{y}_i ,\sigma^2 \right) \text{, where }
	\hat{y}_{it} = \alpha_t + \xi \mu\left(t_i,x_i,\hat{\pi}_i \right) + \left[b_1z_i + b_0(1-z_i) \right]\tau\left(t_i,x_i\right) \\
	%
	p\left(\sigma \middle| \hdots \right) \propto \biggl( p\left(\sigma^2  \right) \cdot p\left(y \middle| \sigma^2 \right) \biggr)
				\sim
	\text{IG}\left( \frac{\nu+n}{2}, \frac{RSS+\nu\lambda}{2}\right) \text{, where }RSS = \sum_{i=1}^{n}\left(y_{it}-\hat{y}_{it} \right)
\end{gather*}
~\\
For updating $\left(\xi \middle| \hdots \right)$,
\begin{gather*}
	\xi \sim N(0,1) \\
	%
	p\left(r_i \middle| \hdots \right) = \prod_{i=1}^{n} \text{N}\left( \underbrace{\frac{y_{i}-\alpha_t - \left[b_1 z_i + b_0 \left(1-z_i \right) \right] \tau(t_i,x_i)}{\mu(t_i,x_i,\hat{\pi}_i)} }_\text{$r_i$}
				\middle| 
				\xi, \frac{\sigma^2}{\mu(t_i,x_i,\hat{\pi}_i)^2} \right) \\
	%
	p\left(\xi \middle| \hdots \right) \propto \biggl( p\left(\xi  \right) \cdot p\left(r_i \middle| \xi \right) \biggr)
	\sim N(\mu^*,v^{2*}) \text{, where }\\
	\begin{align*}
		v^{2*}&=\left(1 + \frac{1}{\sigma^2}\sum_{i=1}^{n}\mu(t_i,x_i,\hat{\pi}_i)^2 \right)^{-1}\\
		\mu^*&= v^{2*}\left(\sum_{i=1}^{n}\mu(t_i,x_i,\hat{\pi}_i)^2r_i  \right)
	\end{align*}
\end{gather*}

~\\
For updating $\left(\Delta_\mu \middle| \hdots \right)$, let $n_\text{bots}$ be the number of leaves $l$ across all $n_\mu$ control trees, and $T$ be the length of the target covariate mesh.  Let $j$ index trees and $l$ index leaves.  The full conditional only uses control tree fits.
\begin{gather*}
	\Delta_\mu \sim Ga\left(\frac{\nu_\mu}{2}, \frac{\nu_\mu}{2} \right) \\
	%
	p\left(m_{jl} \middle| \hdots \right) = \prod_{j,l}^{n_\text{bots}} N_T\left(m_{jl} \mid 0, \frac{s^2_\mu}{\Delta_\mu n_\mu}C_0\right) \text{, where } C_0 = \left[-.5\left(\frac{t-t'}{l_\mu} \right)^2 \right] \\
	%
	p\left(\frac{1}{\Delta_\mu} \middle| \hdots \right) \propto \biggl(p(\Delta_\mu) \cdot p(m_{jl} \mid \Delta_\mu) \biggr)
	\sim IG\left(\frac{\nu_\mu + n_\text{bots}T}{2}, \frac{\nu_\mu + SSQ}{2} \right)
	\text{, where } SSQ=\sum_{j,l}^{n_\text{bots}}m_{jl}^T C_\mu^{-1} m_{jl}
\end{gather*}

~\\
For updating $\left(\Delta_\tau \middle| \hdots \right)$, let $n_\text{bots}$ be the number of leaves $l$ across all $n_\tau$ treatment trees, and $T$ be the length of the target covariate mesh.  Let $j$ index trees, $l$ index leaves.  This full conditional only uses treatment tree fits.
\begin{gather*}
	\Delta_\tau \sim Ga\left(\frac{\nu_\tau}{2}, \frac{\nu_\tau}{2} \right)\\
	%
	p\left(m_{jl} \middle| \hdots \right) = \prod_{j,l}^{n_\text{bots}} N_T\left(m_{jl} \mid 0, \frac{s^2_\tau}{\Delta_\tau n_\tau}C_0\right), \text{ where } C_0 = \left[-.5\left(\frac{t-t'}{l_\tau} \right)^2 \right] \\
	%
	p\left(\frac{1}{\Delta_\tau} \middle| \hdots \right) \propto \biggl(p(\Delta_\tau) \cdot p(m_{jl} \mid \Delta_\tau) \biggr)
	\sim IG\left(\frac{\nu_\tau + n_\text{bots}T}{2}, \frac{\nu_\tau + SSQ}{2} \right)
	\text{, where } SSQ=\sum_{j,l}^{n_\text{bots}}m_{jl}^T C_\tau^{-1} m_{jl}
\end{gather*}

~\\
For updating $\left(b_1 \middle| \hdots \right)$, let $n_z$ be the number of treatment observations.  This full conditional uses only treatment observations.
\begin{gather*}
	b_1 \sim N(\mu_{b_1}=.5, \sigma^2_{b_1}=.5) \\
	%
	p\left(r_i \middle| \hdots \right) = \prod_{i=1}^{n_{z=1}}\text{N}\left( 
				\underbrace{\frac{y_{i}-\alpha_t - \xi \mu(t_i,x_i,\hat{\pi}_i)}{\tau(t_i,x_i)} }_\text{$r_i$}
				\middle| 
				b_1, \frac{\sigma^2}{\tau(t_i,x_i)^2} \right) \\
	%
	p\left(b_1 \middle| \hdots \right) \propto \biggl( p\left(b_1 \right) \cdot p\left(r_i \middle| b_1 \right) \biggr)
	\sim N(\mu^*,v^{2*}) \text{, where }\\
	\begin{align*}
		v^{2*}&=\left( \frac{1}{\sigma^2_{b_1}} + \frac{1}{\sigma^2}\sum_{i=1}^{n_{z=1}}\tau\left(t_i,x_i \right)^2 \right)^{-1}\\
		\mu^*&= v^{2*}\left( \frac{\mu_{b_1}}{\sigma^2_{b_1}} + \frac{1}{\sigma^2}\sum_{i=1}^{n_{z=1}} \tau\left(t_i,x_i \right)^2 r_i \right)
	\end{align*}
\end{gather*}

~\\
For updating $\left(b_0 \middle| \hdots \right)$, let $n_z$ is the number of control observations.  This full conditional uses only control observations.
\begin{gather*}
	b_0 \sim N(\mu_{b_0}=-.5, \sigma^2_{b_0}=.5) \\
	%
	p\left(r_i \middle| \hdots \right) = \prod_{i=1}^{n_{z=0}}\text{N}\left( 
				\underbrace{\frac{y_{i}-\alpha_t - \xi \mu(t_i,x_i,\hat{\pi}_i)}{\tau(t_i,x_i)} }_\text{$r_i$}
				\middle| 
				b_0, \frac{\sigma^2}{\tau(t_i,x_i)^2} \right)\\
	%
	p\left(b_0 \middle| \hdots \right) \propto  \biggl(p\left(b_0 \right) \cdot p\left(r_i \middle| b_0 \right) \biggr)
	\sim N(\mu^*,v^{2*}) \text{, where }\\
	\begin{align*}
		v^{2*}&=\left( \frac{1}{\sigma^2_{b_0}} + \frac{1}{\sigma^2}\sum_{i=1}^{n_{z=0}}\tau\left(t_i,x_i \right)^2 \right)^{-1} \\
		\mu^*&= v^{2*}\left( \frac{\mu_{b_0}}{\sigma^2_{b_0}} + \frac{1}{\sigma^2}\sum_{i=1}^{n_{z=0}} \tau\left(t_i,x_i \right)^2 r_i \right)
	\end{align*}
\end{gather*}

\subsubsection{Marginal likelihood for prognostic tree updates}
The marginal likelihood for updating the prognostic tree fits $\mu(t_i,x_i,\hat{\pi}_i)$ is the version from BART with Targeted Smoothing, with homogeneous variances.  Here, we derive the marginal log-likelihood here for a single leaf.  In the backfitting algorithm, this is calculated for multiple leaves depending on whether a birth move or death move is proposed for the tree.  The likelihoods are then used in calculating the acceptance probability for the Metropolis step.   

Let $y_l$ represent the length $n_l$ vector of residuals for a given leaf.  Let $T_{\mu j}$ be the tree structure for the $j^{th}$ tree, and $t_\text{len}$ be the length of the grid of unique target values.  We integrate out leaf means vector $m_l$ to obtain the marginal log-likelihood as follows.
\begin{align*}
	p(y_l | T_j, \sigma^2) = \int_{\mathbb{R}}
	N_{n_l}\left( y_l \middle|
	W_lm_l, \sigma^2 I\right) \cdot
	N_{t_\text{len}}\left( m_l \middle|
	m_0,K^{-1}
	\right)
	\partial m_l
\end{align*}
where $W_l$ is a $n \times t_\text{len}$ matrix, with one row for each observation; all entries are zero, except a 1 in the column corresponding to each observation $i$'s associated time $t_i$. Set $m_0=0$.  The marginal log-likelihood is then
\begin{align*}
	p(y_l | T_{\mu j}, \sigma^2) = -\frac{n_l}{2}\log\left(2\pi\sigma^2 \right)
	+\frac{1}{2}\log\left(|K| \right)
	-\frac{1}{2}\log\left(|C| \right)
	-\frac{1}{2}\left[ 
		\frac{1}{\sigma^2}y_l^T y_l + m_0^T K m_0 - b^T C^{-1}b 
	\right]
\end{align*}
where $C = \left(\frac{1}{\sigma^2}W_l^TW_l+K\right)$ and $b = \left(\frac{1}{\sigma^2} W_l^T y_l + K m_0 \right)$. For computational purposes, $W_l^TW_l = \bigl[ \begin{smallmatrix} n_1 & \hdots & n_{t_\text{max}} \end{smallmatrix} \bigr]$, the vector of sample sizes for each time.  In addition, $y_l^T y_l = \sum_{i=1}^{n_l}y_i$, the sum of all $y_i$ in leaf $l$.

\subsubsection{Marginal likelihood for treatment tree updates}
The marginal likelihood for updating treatment fits is slightly more complex, requiring heterogeneous variances, since the variance for updating $\tau(t_i,x_i)$ is is $\left(\frac{\sigma^2}{\left[b_1 z_i + b_0 \left(1-z_i \right) \right]^2} \right)$. Let $y_l$ represent the length $n_l$ vector of residuals for a given leaf.  Let $T_{\tau j}$ be the tree structure for the $j^{th}$ tree.  We integrate out leaf means vector $m_l$ to obtain the marginal log-likelihood as follows.  Instead of $\sigma^2I = \left(\omega I \right)^{-1}$, use the $\left(n_l \times n_l \right)$ precision matrix $\Lambda = \text{diag}\left[\omega_1,\hdots, \omega_{n_l}\right]$.

We integrate out leaf means vector $m_l$ to obtain the marginal log-likelihood as follows.
\begin{align*}
	p(y_l | T_{\tau j}, \sigma^2) = \int_{\mathbb{R}}
	N_{n_l}\left( y_l \middle|
	W_l m_l, \Lambda^{-1} \right) \cdot
	N_{T}\left( m_l \middle|
	m_0,K^{-1}
	\right)
	\partial m_l
\end{align*}
where $W_l$ is gain a $n \times t_\text{len}$ matrix, with one row for each observation; all entries are zero, except a 1 in the column corresponding to each observations $i$'s associated time $t_i$.  We again let $m_0=0$.  The marginal log-likelihood is then
\begin{align*}
	p(y_l | T_{\tau j}, \sigma^2) = -\frac{n_l}{2}\log\left(2\pi \right) 
	+ \frac{1}{2} \left[ 
		\log\left(|\Lambda| \right) +
		\log\left(|K| \right) -
		\log\left(|C| \right)
			-\left[y_l^T\Lambda y_l - b^T C^{-1}b \right]
	\right]
\end{align*}
where $C = \left(W_l^T \Lambda W_l+K\right)$ and $b = \left( W_l^T \Lambda y_l + K \mu_0 \right)$.  For computational purposes, $W_l^T \Lambda W_l$ is the the $t_\text{len} \times t_\text{len}$ diagonal matrix of sums of precisions for each sample size.  $W_l^T \Lambda y_l$ is the vector of $\omega_i y_i$ sums for each time point, and $y_l^T \Lambda y_l = \sum_{i=1}^{n_l}\omega_i y_i^2$.  Finally, $\log\left(|\Lambda| \right) = \sum_{i=1}^{n_l} \log\left(\omega_i \right)$

\subsection{Selecting prior scales for marginal tree variances.} \label{appendix-a3}
When modeling a binary response, care must be taken in selecting sensible scale hyperparameters $s_\mu$ and $s_\tau$.  In Section \ref{sec-tsBCF}, we propose an intuitive method for setting these hyperparameters based on a combination of prior elicitation and data-driven specification.  Here, we define structural heterogeneity, provide intuition, make suggestions for setting prior scales, and provide detail to support our finding that approximately 0.02\% of the variance we find in relative risk is comprises structural heterogeneity.

~\\
\textbf{Defining structural heterogeneity.} 
In the scalar response setting, shrinking towards a constant $\tau$ across all observations corresponds to a homogeneous treatment effect.  In the probit case, our causal estimand of interest is relative risk, a non-linear 
transformation which includes both $\mu(t_i,x_i,\hat{\pi}_i)$ and $\tau(t_i,x_i)$. Let $b = \Phi(\alpha+ \mu)$ represent baseline risk, and $r = \frac{\Phi(\alpha + \mu + \tau)}{\Phi(\alpha + \mu)}$ relative risk. If $\tau=0$ for all observations, relative risk is of course constant for all observations.  A fixed non-zero $\tau$ for all observations does not imply homogeneous relative risk, as relative risk will still vary as $\mu(t_i,x_i,\hat{\pi}_i)$ varies.  We call this variability \emph{structural heterogeneity}, and the degree to which it exists depends on the size of $\mu$ and $\tau$, and the size of the model offset $\alpha$.

~\\
\textbf{Setting prior scales.} 
We suggest setting prior scales $s_\mu$ based on prior elicitation of a plausible range of baseline risks, and $s_\tau$ using estimation of baseline risk and relative risk from a small amount of held-out data.  We select a range of 0.86 to 0.999 as a plausible range of relative risks, and estimate that the range of plausible values covers about 3.3 times the standard deviation, such that
\begin{align*}
	s_\mu = \frac{\Phi^{-1}\left(.999 \right) - \Phi^{-1}\left(.860 \right)}{3.3}.
\end{align*}
We then hold out 500 observations, obtain point estimates for relative risk and baseline risk, and use Nelder Mead to solve for a reasonable $s_\tau$ value.

\begin{figure}[ht]
		\centering
			\includegraphics[width=0.85\textwidth]{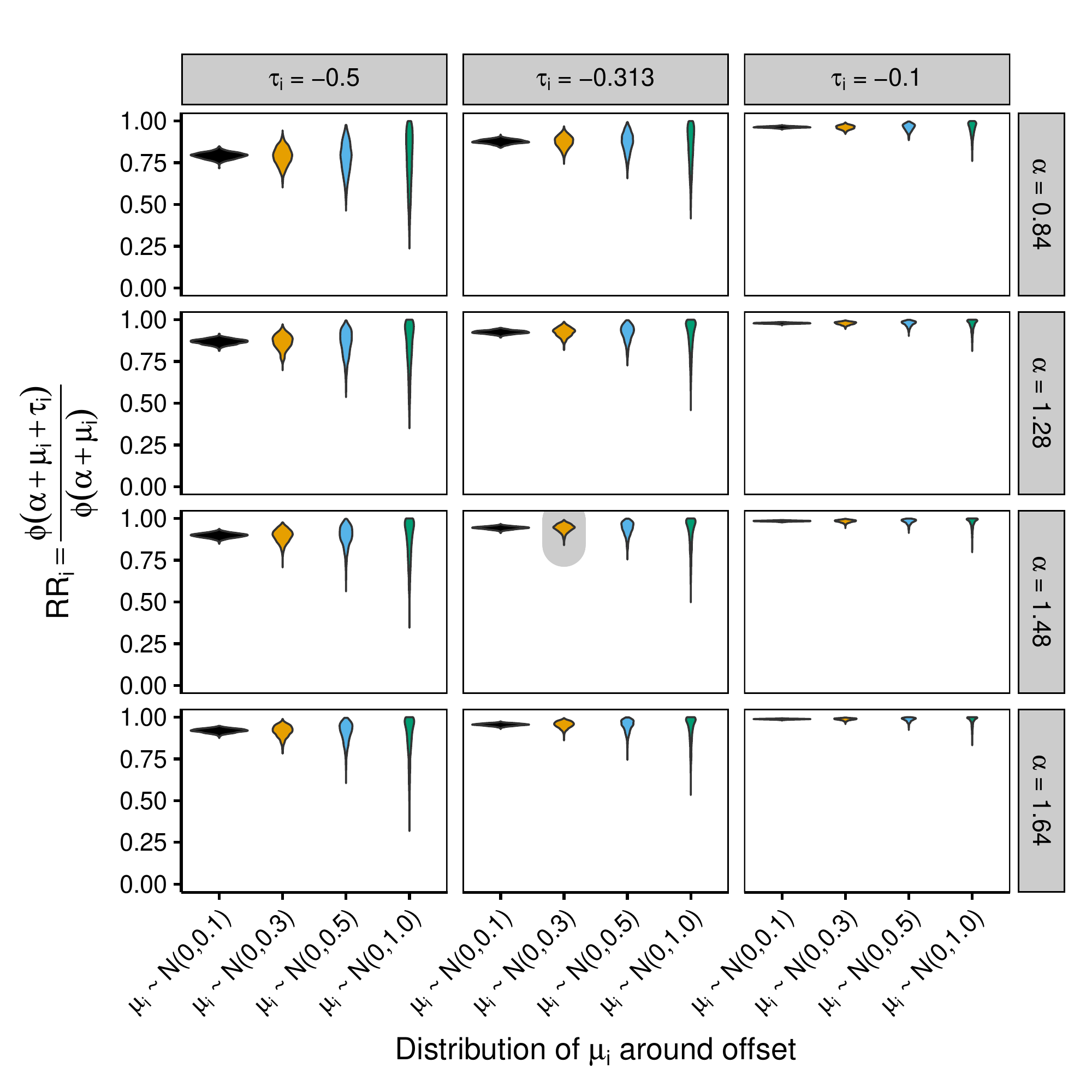}
		\caption{Illustrates structural heterogeneity in relative risk for six combinations of offset $\alpha$ and treatment effect $\tau_i$, with $\mu_i$ coming from distributions with different levels of variability.  Amount of structural heterogeneity depends on the size of the offset, the magnitude of $\tau_i$, and the variability in $\mu_i$; structural heterogeneity generally increases as $\tau_i$ increases in magnitude and as $\mu_i$ is more disperse, and generally decreases as baseline risk increases. The violin plot highlighted in grey ($\alpha$=1.48, $\tau_i$=-0.313, and $\mu_i \sim N(0,0.3)$) corresponds most closely to our early medical abortion analysis; this combination corresponds to our average offset, posterior mean $\tau$, and standard error of the posterior $\mu_i$ estimates from our analysis.}
		\label{fig:prob-ill}
	\end{figure}

~\\
\textbf{Structural heterogeneity intuition.} 
Figure \ref{fig:prob-ill} provides intuition about the behavior of structural heterogeneity.  Each row represents a different offset ($\alpha$), corresponding to average baseline risks of 0.80, 0.90, 0.93, and 0.95; columns show three different $\tau_i$ values, $-0.5$m $-0.313$, and $-0.1$.  For each of the $(\alpha, \tau_i)$ combinations, we draw 1,000 $\mu_i$ from four Gaussian distributions with increasing variance and calculate the individual relative risks $RR_i = \frac{\Phi\left(\alpha + \mu_i + \tau_i \right)}{\Phi\left(\alpha + \mu_i \right)}$.  Amount of structural heterogeneity depends on the size of the offset, the magnitude of $\tau_i$, and the variability in $\mu_i$; structural heterogeneity generally increases as $\tau_i$ increases in magnitude and as $\mu_i$ is more disperse, and generally decreases as baseline risk increases.  The violin plot highlighted in grey, where $\alpha$=1.48, $\tau_i$=-0.313, and $\mu_i \sim N(0,0.3)$ corresponds most closely to our early medical abortion analysis; this combination corresponds to the average offset, the posterior mean $\tau$ value, and the standard error of the posterior $\mu_i$ estimates from our analysis.

~\\
\textbf{Quantifying structural heterogeneity.} 
We estimate the degree to which structural heterogeneity is present in our early medical abortion analysis as follows.  We set prior scales $s_\mu$, $s_\tau$ as described above.  For each posterior draw $b$, we calculate the variance in individual relative risk as
\begin{align} \label{eqn:rr-var}
	\hat{\sigma}_{RR}^{(b)} = \text{Var}\left(\widehat{RR}_i^{(b)} \right)
\end{align}
where $RR_i^{(b)}$ (\ref{eqn:relrisk}) is estimated using posterior draws $\mu_i^{(b)}$, $\tau_i^{(b)}$.  

Let $\bar{\tau}^{(b)} = \sum_{i=1}^{n}\tau_i^{(b)}$ represent the mean of $\tau$ across all observations for draw $b$.  We now let $\hat{\sigma}_{RR_{het}}^{(b)}$ represent a second estimate of the variance in $RR_i^{(b)}$ for draw $b$, calculated as in (\ref{eqn:rr-var}) but replacing posterior draws $\tau_i^{(b)}$ with heterogeneous $\bar{\tau}^{(b)}$.

We estimate how much more variability is present in relative risk under heterogeneous $\tau_i$ compared to homogeneous $\tau$ by computing the ratio of $\hat{\sigma}_{RR}^{(b)}$ to $\hat{\sigma}_{RR_{het}}^{(b)}$ for each draw and averaging across draws.  
We find that there is an average of 2.06 times more variability in posterior relative risk under heterogeneous versus homogeneous $\tau_i$, giving us confidence that our clinical findings are not just an artifact of structural heterogeneity.

\end{document}